\documentclass[11pt,notitlepage]{article}
\usepackage{graphicx}
\usepackage{dcolumn}
\usepackage{amsmath, amsthm, amssymb, esint}
\usepackage{longtable}
\usepackage{fullpage}
\usepackage{upgreek}
\usepackage{comment}
\usepackage{multirow}
\usepackage{overpic}
\usepackage{authblk}

\usepackage[numbers,sort&compress]{natbib}

\usepackage{gensymb}

\newcommand{\diff}{\mathrm{d}} 

\begin{document}

\title{Principles and Techniques of the Quantum Diamond Microscope}

\author[1, 2]{\small Edlyn V. Levine \footnote{These authors contributed equally to this work.}\footnote{Correspondence to evlevine@mitre.org}}
\author[1, 3]{\small Matthew J. Turner*}
\author[4]{\small Pauli Kehayias}
\author[1]{\small Connor A. Hart}
\author[1,5]{\small Nicholas Langellier}
\author[1,6]{\small Raisa Trubko}
\author[1]{\small David R. Glenn}
\author[6]{\small Roger R. Fu}
\author[5]{\small Ronald L. Walsworth}

\affil[1]{\footnotesize Department of Physics, Harvard University, Cambridge, MA 02138, USA}
\affil[2]{\footnotesize The MITRE Corporation, Bedford, MA 01730, USA}
\affil[3]{\footnotesize Center for Brain Science, Harvard University, Cambridge, MA 02138, USA}
\affil[4]{\footnotesize Sandia National Laboratories, Albuquerque, NM 87123, USA}
\affil[5]{\footnotesize Harvard-Smithsonian Center for Astrophysics, Cambridge, MA 02138, USA}
\affil[6]{\footnotesize Department of Earth and Planetary Sciences, Harvard University, Cambridge, MA 02138, USA}
\affil[*]{\footnotesize E.V.L and M.J.T contributed equally to this work.}
\affil[$\dagger$]{\footnotesize Correspondence to evlevine@mitre.org}

\maketitle

\begin{abstract}
    We provide an overview of the experimental techniques, measurement modalities, and diverse applications of the Quantum Diamond Microscope (QDM).  The QDM employs a dense layer of fluorescent nitrogen-vacancy (NV) color centers near the surface of a transparent diamond chip on which a sample of interest is placed.  NV electronic spins are coherently probed with microwaves and optically initialized and read out to provide spatially resolved maps of local magnetic fields.  NV fluorescence is measured simultaneously across the diamond surface, resulting in a wide-field, two-dimensional magnetic field image with adjustable spatial pixel size set by the parameters of the imaging system.  NV measurement protocols are tailored for imaging of broadband and narrowband fields, from DC to GHz frequencies. Here we summarize the physical principles common to diverse implementations of the QDM and review example applications of the technology in geoscience, biology, and materials science. 
\end{abstract}

\section{Introduction}
Nitrogen-vacancy (NV) centers in diamond are a leading modality for sensitive, high-spatial-resolution, wide field of view imaging of microscopic magnetic fields. NV-diamond sensors operate in a wide range of conditions, from cryogenic to well above room temperature, and can serve as broadband detectors of slowly-varying magnetic fields, or as narrow-band detectors of magnetic fields over a wide range of frequencies from near DC to GHz. Full vector magnetic field sensing is possible using the distribution of NV orientations along the four crystallographic directions in diamond. 

NV centers function at ambient conditions, and have magnetically, electrically, and thermally sensitive electronic spin ground states with long coherence lifetimes. The NV spin state can be initialized, and the evolution of the spin states can be detected optically, thus allowing precision sensing of magnetic fields and other effects. Magnetic field sensitivity and spatial resolution are determined by the number of NVs in the sensing volume, the resonance linewidth, the resonance spin-state fluorescence contrast, the collected NV fluorescence intensity, and the NV-to-sample separation. 

Variation of the experimental setup and measurement protocol allows NV-diamond magnetic imaging to be adapted for a wide range of applications in different fields of research. Although the desired capabilities for each magnetic imaging application vary widely, common requirements include good field sensitivity within a defined frequency range, fine spatial resolution, large field of view, quantitative vector magnetometry, wide field and frequency dynamic range, and flexibility in the bias field and temperature during measurement. For example, imaging for geoscience \cite{QDM1ggg} and cell biology \cite{degenReview} applications generally require high sensitivity to DC magnetic fields, spatial resolution at the optical diffraction limit, and room temperature operation. In contrast, microelectronics magnetic field imaging \cite{ICGHz} can require magnetic field frequency sensitivity up to the GHz range. Applications that do not require simultaneous imaging over a wide-field of view can also leverage scanning magnetometry using single NV centers at the tips of monolithic diamond nanopillars, or in nanodiamonds at the ends of atomic force microscopy cantilevers \cite{degen:243111,rondin_imaging1,afm3}. 

With proper optimization, NV diamond magnetometry can offer combinations of the above capabilities unattainable using alternative magnetic imaging techniques. The Magnetic Force Microscope (MFM) \cite{MFMReview}, while offering higher spatial resolution, is limited by small ($<$100 $\upmu$m) fields of view, worse DC field resolution ($>$10 $\upmu$T), and potential complications due to sensor-sample interactions. Meanwhile, the SQUID microscope, if measuring a room-temperature sample, can only achieve spatial resolution of $>$150 $\upmu$m, although with excellent DC sensitivity ($<$500 fT/$\sqrt{\text{Hz}}$) \cite{FONGSQUID}. Finally, other techniques such as Magneto-Optical Kerr Effect (MOKE) \cite{MOKE1, MOKE2} and other Faraday Effect-based magneto-optical imaging cannot produce reliable, quantitative maps of the vector magnetic field.  

\begin{figure}[ht]
\begin{center}
\includegraphics[width=0.75\textwidth]{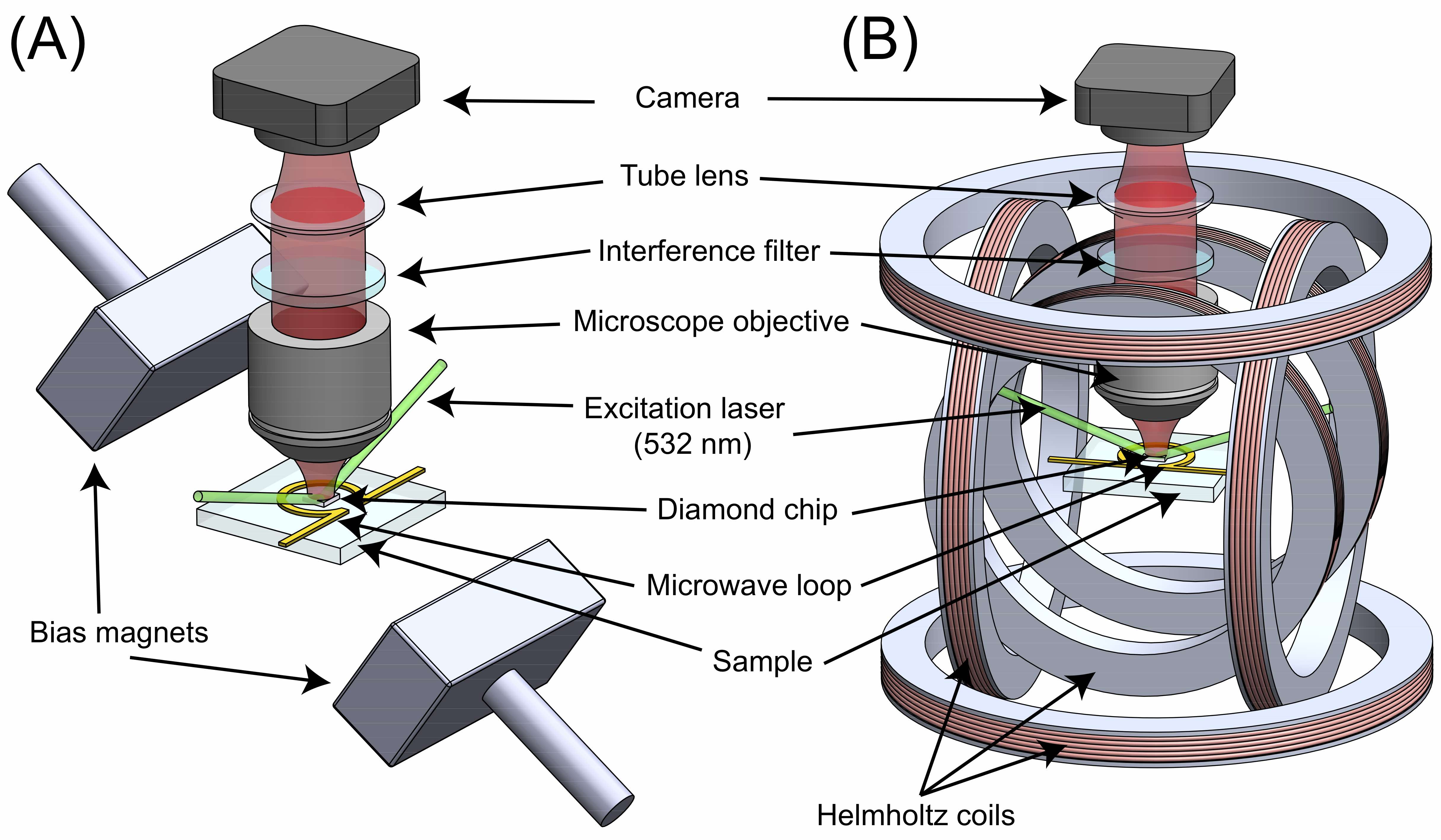}
\end{center}
\caption{Quantum Diamond Microscope (QDM). Examples with (A) permanent magnets and (B) Helmholtz coils to apply a bias magnetic field. In both configurations, 532 nm excitation laser light illuminates the diamond chip and optics collect NV fluorescence onto a camera. The interference filter is chosen to transmit NV fluorescence and, in particular, block scattered excitation light. A planar, gold omega-loop, fabricated onto a substrate, is depicted delivering microwaves to the diamond chip for NV control.
}. \label{Fig1}
\end{figure}

This review article provides an overview of the Quantum Diamond Microscope (QDM), a common approach to ensemble NV wide-field magnetic imaging, and describes specific optimization of the QDM for several applications \cite{QDM1ggg, nv_bacteria}.  Schematics of typical QDM setups are shown in Fig.~\ref{Fig1}. The QDM uses an optical microscope and a camera to measure the fluorescence from a thin ensemble NV layer at the surface of the diamond sensor chip, with the sample placed near to or in contact with the diamond. The local magnetic field of the sample is extracted from each camera pixel measurement and a wide-field map of the magnetic field is constructed from the pixel array. We present the methods needed to image static and dynamic magnetic fields with the QDM, and briefly discuss imaging of temperature and electric fields. For each type of sample field -- narrow-band, broadband, etc. -- we describe the quantum control procedures and hardware choices that are necessary for ideal imaging, and emphasize the design trade-offs in optimal sensitivity and resolution limits that can be achieved.

The article is organized as follows:

\tableofcontents{}

\section{NV Physics Relevant to QDMs}
QDM implementation, including assembly and method of operation, depends on the intended application and the characteristics of the sample fields. However, there are principles of NV physics relevant to all QDM experiments. These principles rely on single NV spin properties and their ensemble behavior. The QDM has optical, static magnetic, and microwave (MW) fields that are applied to  manipulate the NV electronic- and spin-state populations in a controlled manner. An unknown sample field modifies the NV spin states, and is detectable by changes in NV fluorescence. The three QDM driving fields are chosen to optimize coupling between the sample field and the NV spin state.  

\subsection{NV Ground Electronic State in the Absence of External Fields}
\begin{figure}[ht]
\begin{center}
\begin{overpic}[width=\columnwidth]{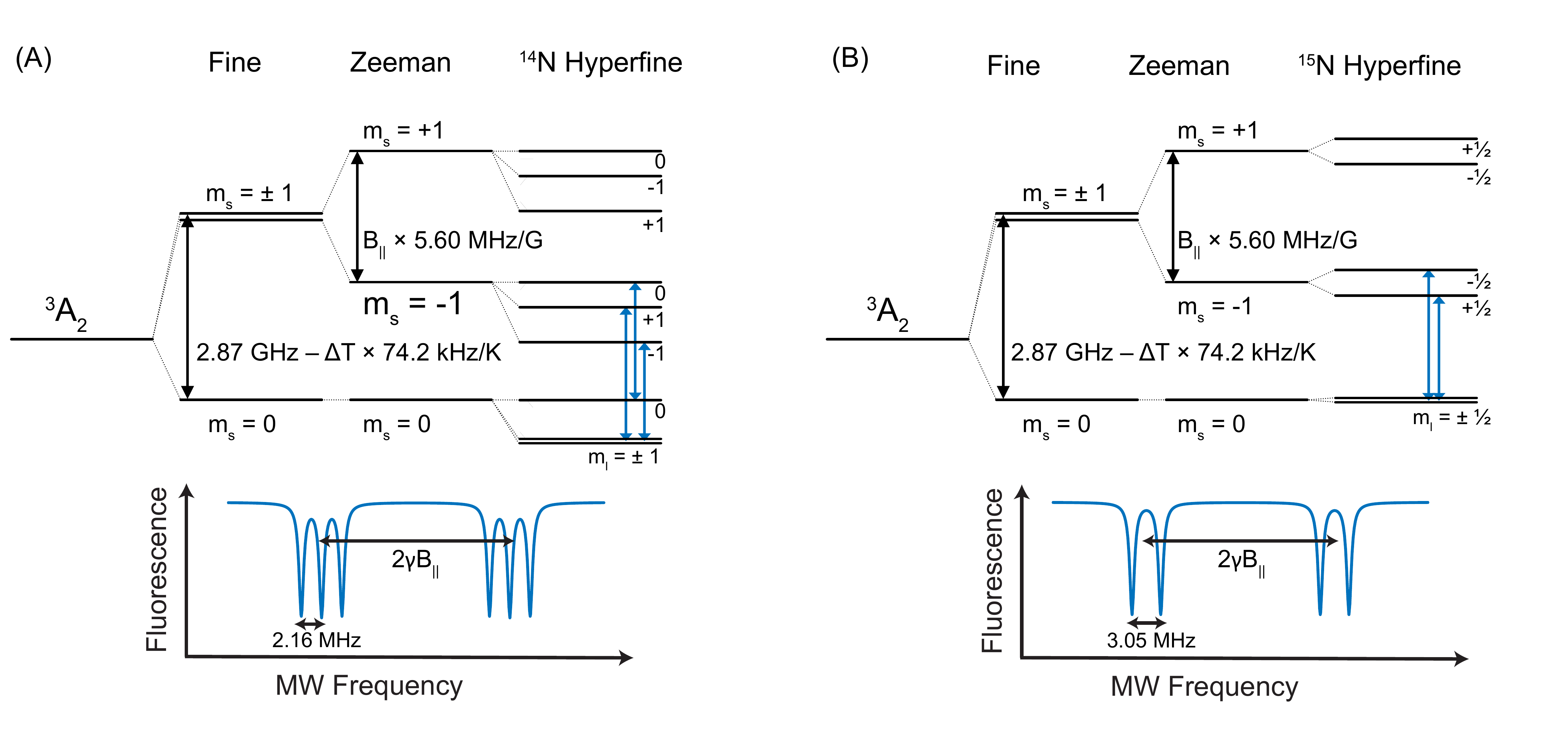}
\end{overpic}
\end{center}
\caption{\label{3A2energies}
NV Ground State Configurations and ODMR Spectra. (A) $^{14}$N hyperfine states and
(B) $^{15}$N hyperfine  states. Schematic optically-detected magnetic resonance (ODMR) spectra are shown with Zeeman splitting and hyperfine splitting for $^{14}$N and $^{15}$N. The energy levels for $^{14}$N are further shifted by quadrupolar interactions.}
\end{figure}

Quantum control of NV centers with the QDM driving fields is possible due to the NV electronic and spin level structure \cite{marcus_review,frMagReview}. An NV center consists of a substitutional nitrogen and an adjacent lattice vacancy defect in a diamond crystal. A negatively-charged NV has six electrons, with two electrons from nitrogen, one electron from each of the three carbon atoms, and an additional electron from the lattice. These electrons occupy four $sp^3$ atomic orbitals with electronic spin quantum number $S=1$. These $sp^3$ orbitals linearly combine to form four molecular orbitals \cite{alp_ISC} comprising the ground electronic configuration. The lowest energy state of the ground configuration is the orbital singlet, spin triplet state, $^3A_2$ which has fine, Zeeman, and hyperfine structures shown in Fig.~\ref{3A2energies}. The four molecular orbitals also give rise to electronic excited states: orbital-doublet spin-triplet $^3E$, and spin-singlet orbital-singlet $^1A_1$ shown in Figure \ref{allEnergies}.

NV magnetometry uses fluorescence from electronic state transitions to detect changes to the $^3A_2$ ground state configuration that result from coupling to a sample field. Therefore, focus is placed on the physics of the $^3A_2$ Hamiltonian. NV centers have $C_{3v}$ point-group symmetry, and are spatially invariant under the $C_{3v}$ symmetry transformations (the identity, two 120$^\circ$ rotations about the N-V axis, and three vertical reflection planes). 

NV centers also have a built-in quantization axis along the NV axis (called the NV $z$-axis, or the crystallographic [\textit{111}] direction). The $^3A_2$ electronic ground state is an orbital singlet and spin triplet manifold, with ground-state Hamiltonian \cite{LoubserVanWyk1978}
\begin{equation}
\frac{\hat{H}_{gs}}{h} = \hat{S}\cdot \mathbf{D} \cdot\hat{S}+\hat{S}\cdot \mathbf{A} \cdot\hat{I} +\hat{I}\cdot \mathbf{Q} \cdot\hat{I},
\end{equation}
where $h$ is Planck's constant and $\hat{S}= (\hat{S}_x, \hat{S}_y, \hat{S}_x)$ and $\hat{I}= (\hat{I}_x, \hat{I}_y, \hat{I}_x)$ are the dimensionless electron and nitrogen nuclear spin operators, respectively. The first term is the fine structure splitting due to the electronic spin-spin interaction, with the fine structure tensor $\mathbf{D}$ \cite{marcusGndState}. The second term is the hyperfine interaction between NV electrons and the nitrogen nucleus with the hyperfine tensor $\mathbf{A}$. The third term is the nuclear electric quadrupole interaction, with electric quadrupole tensor $\mathbf{Q}$. Under the $C_{3v}$ symmetry of the NV center, $\mathbf{D}$, $\mathbf{A}$, and $\mathbf{Q}$, are diagonal in the NV coordinate system \cite{P1DQ, PhysRevB.89.205202}, and $\hat{H}_{gs}$ can be written as \cite{marcus_review}
\begin{equation}
    \frac{\hat{H}_{gs}}{h} = D(T)\left[\hat{S}_z^2 -\hat{S}^2/3\right]+ A^\parallel \hat{S}_z\hat{I}_z +A^\perp\left[\hat{S}_x\hat{I}_x+\hat{S}_y\hat{I}_y\right]+P\left[\hat{I}_z^2-\hat{I}^2/3\right]
    \label{hfTerms}.
\end{equation}
$D(T)$ is the fine structure term called the zero-field splitting (ZFS), $A^\parallel$ and $A^\perp$ are the axial and transverse hyperfine terms, and $P$ is the nuclear electric quadrupole component. Two important features of the ground state are evident from the Hamiltonian. First, the $^3A_2$ $m_s = \pm 1$ magnetic sublevels and the $m_s = 0$ have $D(T)$ difference in energy. $D(T)$ is temperature dependent due to spin-spin interaction changes with the lattice constant \cite{victor_Tdepend, marcus_temperature}, with $D \sim 2.87$ GHz and $dD/dT = -74.2$ kHz/K at room temperature. Second, the $^3A_2$ electronic states have an additional hyperfine energy splitting $A_{gs}^\parallel \hat{S}_z\hat{I}_z$  due to the nitrogen nucleus. $I=1$ for a $^{14}$N nucleus while $I=1/2$ for a $^{15}$N nucleus. The energy level diagrams for $^{14}$N and $^{15}$N are shown in Fig.~\ref{3A2energies}. Hyperfine parameters are $A_{14N}^\parallel = -2.14$ MHz, $A_{14N}^\perp = -2.70$ MHz, $P_{14N}^\parallel = -5.01$ MHz, $A_{15N}^\parallel = 3.03$ MHz, $A_{15N}^\perp = 3.65$ MHz \cite{felton_hf}.

Crystal stress in the diamond also contributes to the $^3A_2$ Hamiltonian. This is expressed as \cite{marcusStrainHam, galiSpinStrain, maletinskyStrainTerms},
\begin{equation}
    \frac{\hat{V}_{str}}{h} = M_{z}\hat{S}_z^2+ M_{x}(\hat{S}^2_x-\hat{S}^2_y) + M_{y}(\hat{S}_x\hat{S}_y+\hat{S}_y\hat{S}_x)+N_x(\hat{S}_x\hat{S}_z+\hat{S}_z\hat{S}_x)+N_y(\hat{S}_y\hat{S}_z+\hat{S}_z\hat{S}_y).
    \label{strain}
\end{equation}
Here $M_{x}$, $M_{y}$, $M_{z}$, $N_{x}$, and $N_{y}$ are stress-dependent amplitudes. The $M_{z}$ term contributes to the zero-field splitting, while the other terms may be negligible or suppressed depending on the experimental conditions (such as applied magnetic field \cite{EBnoiseCompetition}). The NV spin sensitivity to this spin-stress induced interaction can be used to image internal or external diamond stress \cite{strainMapping}, which is important for diamond material characterization. However, for imaging external magnetic fields, we consider NV stress sensitivity as a potential limitation.

\subsection{NV Electronic Transitions}\label{NVStates}
\begin{figure}[ht]
\begin{center}
\includegraphics[width=0.5\textwidth]{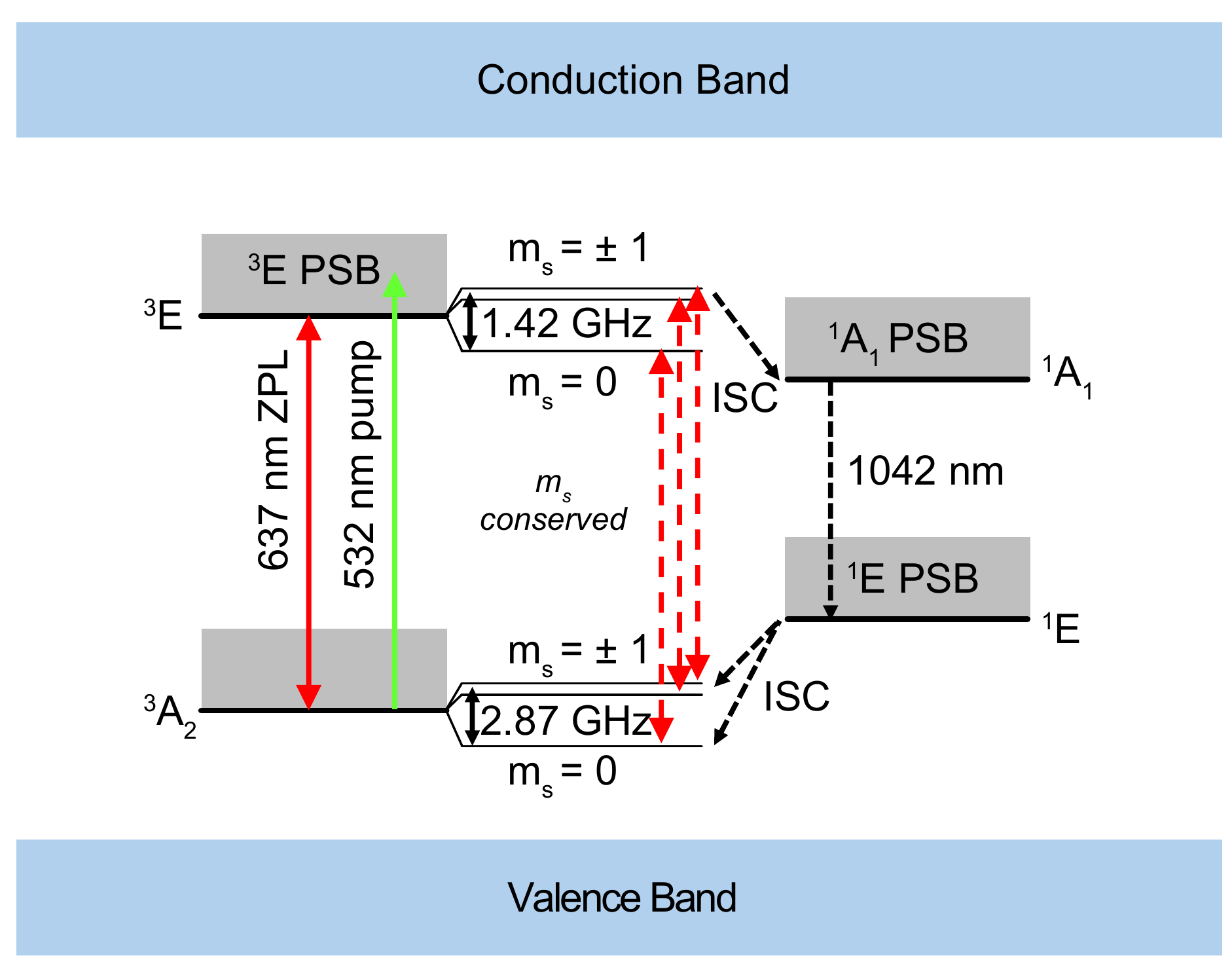}
\end{center}
\caption{\label{allEnergies} NV Radiative and Non-radiative State Transitions. Radiative $^3E\leftrightarrow ^3A_2$ transition with optical 637nm zero-phonon line (ZPL), and $^1E \leftrightarrow {^1}A_1$ transition with non-optical 1042 nm ZPL. Phonon sidebands (PSBs) can shift the transition frequencies. Non-radiative intersystem crossing (ISC) mediated transitions exist between $^3E$ and $^1A_1$, and $^1E$ and $^3A_2$.
}
\end{figure}

The NV first excited electronic configuration has an orbital-doublet, spin-triplet state, $^3E$ shown in Fig.~\ref{allEnergies}. The two orbital states and three spin states of $^3E$ combine to form six fine structure states that reduce to three states at room temperature \cite{1367-2630-11-6-063007}, resembling the $^3A_2$ state. $^3E$ is coupled to the $^3A_2$ ground state by an optical 637 nm zero-phonon line (ZPL). The $^3E \leftrightarrow {^3}A_2$ is a radiative transition that generally conserves the electron spin state $m_s$ as a result of weak spin-orbit interaction \cite{3Espec3}. The $^3E \rightarrow ^3A_2$ ($^3A_2 \rightarrow ^3E$ ) transition works for longer [shorter] wavelengths in fluorescence [absorption] as a result of the phonon sideband (PSB). This behavior is similar to Stokes and anti-Stokes shifted transitions \cite{henderson_imbusch}. Fig.~\ref{allEnergies} also shows the radiative, spin conserving $^1E \leftrightarrow {^1}A_1$ transition which has an infrared 1042 nm ZPL and its own sideband structure.

Nonradiative transitions between states of different spin multiplicity exist between $^3E$ and $^1A_1$, and between $^1E$ and $^3A_2$. These nonradiative transitions are caused by an electron-phonon mediated intersystem crossing (ISC) mechanism, and do not conserve spin. The probability of the ISC transition occurring for the $^3E$ to $^1A_1$ is only non-negligible for $m_s=\pm1$ states of $^3E$ and is characterized by the ISC rate of transition \cite{alp_ISC}. Similarly,  the ISC transition probability from $^1E$ to the $m_s=0$ state of $^3A_2$ is approximately 1.1 to 2 times that of the ISC transition from $^1E$ to the $m_s=\pm1$ states of $^3A_2$ \cite{robledo_rates, tetienne_rates}. These state-selective differences in the ISC transition rate allow for spin polarization of the NV under optical excitation from 532 nm laser illumination.

\subsection{Optical Pumping and Spin Polarization}
An optical driving field from a pump laser is applied in order to spin polarize the NV electronic state at the start of a QDM measurement. This pump laser is also used at the end of a measurement to read out the final NV spin state through the fluorescence intensity. NV optical pumping takes advantage of the $m_s$-selective nonradiative ISC decay pathway \cite{robledo_rates, tetienne_rates}. An NV that is optically excited from $^3A_2$ to  $^3E$ state by a 532 nm photon, decays along either the optically radiative $^3E \rightarrow {^3}A_2$ pathway or the non-optical, ISC mediated $^3E \rightarrow ^1A_1 \rightarrow ^1E \rightarrow {^3}A_2$ pathway. The $m_s$-selectivity of the ISC will preferentially depopulate the $m_s=\pm1$ spin projection states. NVs starting in the $^3A_2$ $m_s=\pm1$ sublevel are eventually pumped (on average, after a few pump photon absorption cycles) into the $^3A_2$ $m_s=0$ sublevel.  Typically only $\sim 80\%$ of NVs in an ensemble can be initialized into the $m_s=0$ state \cite{HARRISON2006586}, where they remain in a cycling transition. The ${^1}E$ state is metastable with a $\sim$200 ns lifetime at room temperature \cite{manson_singlets, victor_singlets}. The $^3E$ upper state has a $t_{3E} \approx 13 $ ns lifetime \cite{robledo_rates, tetienne_rates}, and the $^3A_2 \rightarrow {^3}E$ absorption cross section at $\lambda = 532$ nm \cite{wee_2photon} is $\sigma = 3.1\times10^{-17}$ cm$^2$ (although there is disparity in the reported 532 nm absorption cross section value and saturation intensity \cite{ChapmanCrossSection, loncarCrossSec}). These corresponds to a $ (h c)/(\lambda \sigma t_{3E}) \approx 0.9$ MW/cm$^2$ saturation intensity, where $c$ is the speed of light. 

The ISC is also responsible for the reduced fluorescence intensities of NVs in the $m_s=\pm1$ sublevels, since they emit fewer optical photons when returning to the $^3A_2$ state through the ISC mediated pathway. The fractional fluorescence difference between NVs in the $m_s=\pm1$ sublevels and NVs in the $m_s=0$ sublevel is called the fluorescence contrast, and can be as large as $\sim$20$\%$ for a single NV \cite{dreau_powBroad}. The fluorescence intensity from an optically pumped NV diamond chip therefore indicates the percentage of the NVs in the $m_s=0$ state, or in the $m_s=\pm 1$ states. A transition of NVs from the $m_s=0$ to the $m_s=\pm 1$ state, e.g. induced by a resonant MW field, drops the fluorescence as more NVs follow the ISC mediated decay transition. This is the mechanism underlying optical readout for QDM imaging. 

\subsection{Microwave Driving Field}
A MW driving field resonant with the $m_s = 0$ to $+1$ or $-1$ transitions induces Rabi oscillations, transferring the NV population from one sublevel to the other, and creating superpositions of $m_s$ states. Either a continuous-wave (CW) or a pulsed MW field can be used. The length of the MW pulse determines its impact on the NV population: $\pi$ pulses are of sufficient duration to transfer the NV population from the $m_s=0$ to $m_s=1$ when the NVs are initialized in the $m_s=0$ state; $\pi/2$ pulses are of duration to create an equal superposition of $m_s$ states. The utility of pulsed MW fields for QDM detection of different types of sample fields will be discussed below.  

Applying resonant CW MWs simultaneously while optically pumping of the NVs to the $m_s=0$ sublevel results in MW-induced transfer of the NV population out of the $m_s=0$ sublevel, spoiling the optical spin polarization and decreasing the emitted fluorescence intensity. Measuring the NV fluorescence intensity as a function of the probing MW frequency is called Optically-Detected Magnetic Resonance (ODMR) spectroscopy \cite{Gruber}. Simulations of ensemble NV ODMR spectra for NVs with $^{14}$N and $^{15}$N isotopes are shown in Fig.~\ref{3A2energies}. The known dependencies of the $^3A_2$ sublevel energy on external fields allow conversion of these ODMR spectra into magnetic field, electric field, temperature, and crystal stress information. 

\subsection{Static Magnetic Bias Field and Zeeman Splitting}
A static magnetic field $\mathbf{B}_0$ causes a Zeeman interaction in the $^3A_2$ state, written as 
\begin{equation}
    \frac{\hat{V}_{mag}}{h} = \frac{\mu_B}{h} \mathbf{B}_0\cdot\mathbf{g}\cdot\hat{S}= \frac{g_e\mu_B}{h}(B_{0x}\hat{S}_x+B_{0y}\hat{S}_y+ B_{0z}\hat{S}_z).
    \label{zeeman}
\end{equation}
Here, $\mu_B = 9.27\times10^{-24}$ J/T is the Bohr magneton, $\mathbf{g}$ is the electronic g-factor tensor (which is nearly diagonal), $g_e\approx 2.003$ equal to the NV center's electronic g-factor \cite{marcus_review}, and $\gamma=g_e\mu_B/h$ is the NV gyromagnetic ratio. The Zeeman interaction lifts the degeneracy between the $m_s=\pm1$ sublevels, and for $|\mathbf{B}_0|$ along the N-V axis, the $m_s=\pm1$ sublevel energies split linearly with $|\mathbf{B}_0|$ while the $m_s=0$ sublevel is unaffected. The nuclear Zeeman terms are considered negligible and have been excluded.

A sufficiently large bias magnetic field makes the Zeeman term dominant in the Hamiltonian. Otherwise, terms including stress and electric field would dominate with the Zeeman term acting as a perturbation, reducing magnetic field sensitivity and complicating the data analysis. For magnetic imaging, both the static bias fields, which are part of the QDM, and the sample magnetic fields, contribute to the Zeeman interaction.

\subsection{Sample Fields}
QDM experiments create a two dimensional image of the magnetic fields from a sample containing a distribution of magnetic field sources. It is also possible to image a sample's temperature distribution and electric fields. 

\begin{figure}[ht]
\begin{center}\includegraphics[width=1.0\textwidth]{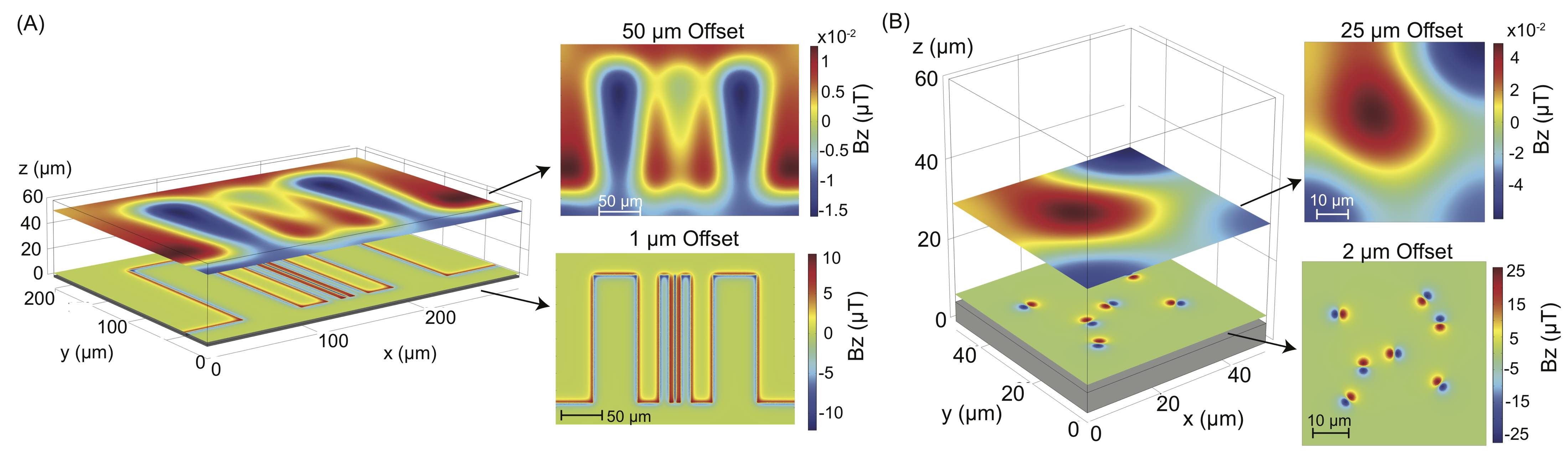}
\end{center}
\caption{\label{BFields} Simulated QDM Measurement Planes above Magnetic Samples. 
Magnetic field distributions from (A) current distributions and (B) magnetic dipole distributions simulated in COMSOL. The NV layer in the QDM measures the sample magnetic field in the $x$-$y$ plane a distance $z$ above the sample. Two measurement planes at different values of $z$ are shown for each simulation. A smaller stand-off distance between the measurement plane and the sample gives a magnetic field image with higher spatial resolution. 
}
\end{figure}

The sample magnetic field is generated by field sources, such as current densities or magnetic dipoles, with either known or unknown distributions. Measurement of the sample magnetic field can be used for the inverse problem of estimating an unknown source distribution under certain conditions \cite{doi:10.1063/1.342549,LimaInverse,MomentEstimation}. The form of the sample magnetic field in terms of its sources is 
\begin{align}
  \begin{aligned}
    &\mathbf{B}_{s}(\mathbf{r},t)=\,\frac{\mu_0}{4\pi}\,\iiint\diff V' \left[\mathbf{J}(\mathbf{r}',t)\times\frac{ \mathbf{\hat{R}}}{R^2}   \right]
  \end{aligned}
  \label{BiotSavart}
\end{align}
where $\mathbf{J}$ is the current density of the sample, $\mathbf{R}=\mathbf{r}-\mathbf{r}'$ is the distance from a magnetic source at $\mathbf{r}'$ to an observation point at $\mathbf{r}$, $R=|\mathbf{r}-\mathbf{r}'|$, and $ \mathbf{\hat{R}}=\mathbf{R}/R$. Eqn.~\ref{BiotSavart} is the Biot-Savart Law for static fields, and applies to fields in the quasistatic regime for which the characteristic system size is small compared to the electromagnetic wavelength. A sample consisting of small magnetic particles will have a magnetic field composed of single dipole fields of the form
\begin{align}
  \begin{aligned}
    &\mathbf{B}_{s}(\mathbf{r})=\,\frac{\mu_0}{4\pi}\left[\frac{3\mathbf{n}(\mathbf{n}\cdot\mathbf{m})-\mathbf{m}}{r^3}+\frac{8\pi}{3}\mathbf{m}\,\delta(\mathbf{r}) \right]
  \end{aligned}
  \label{DipoleField}
\end{align}
Here $\mathbf{m}$ is the magnetic moment, $\mathbf{n}=\mathbf{r}/r$, and the delta function only contributes to the field at the site of the dipole $\mathbf{r}=0$. Other typical sample fields, such as the narrowband magnetic field from the Larmor precession of protons can also be derived \cite{Glenn2018}. Fig.~\ref{BFields} shows examples of the magnetic fields for a current distribution and a distribution of magnetic dipoles. 

The time dependence of the sample magnetic field will determine the QDM measurement protocol. Static and quasi-static sample fields will contribute to the NV Hamiltonian by an additional term in Eqn.~\ref{zeeman}
\begin{equation}
    \frac{\hat{V}_{mag}}{h} = \frac{g_e\mu_B}{h}(\mathbf{B}_0+\mathbf{B}_s)\cdot\hat{S}
    \label{Bsample}
\end{equation}
were $\mathbf{B}_s$ is the magnetic field of the sample, which can take the forms given in Eqns.~\ref{BiotSavart} and \ref{DipoleField}. The magnitude of the sample field along the NV axis is therefore determined by changes in separation in the ODMR resonance features that result from $\mathbf{B}_s$ in addition to the effect of the bias field. ODMR measurements, with and without the sample, then allow determination of the unknown $\mathbf{B}_s$ field. Reasonable assumptions can be made to determine $B_s$ without having to take multiple measurements \cite{QDM1ggg}. A time-varying sample magnetic field with frequency components near the 2.87 GHz ZFS will in turn induce NV spin transitions if the $\mathbf{B}_0$ bias field has been tuned to the appropriate Zeeman splitting. Sweeping the $\mathbf{B}_0$ field will then locate the frequency of the sample fields, with magnitude determined by the ODMR contrast depth and line width. 

Electric field and temperature distributions from the sample will also change the NV spin states. The external sample electric field $\mathbf{E}_s = (E_{sx},E_{sy},E_{sz})$ adds to the internal local electric fields \cite{PhysRevLett.121.246402}, $\mathbf{E}_{loc} = (E_{loc,x},E_{loc,y},E_{loc,z})$ in the diamond, e.g. induced by a high density of P1 (Nitrogen) centers, such that $\mathbf{E}_{tot} =\mathbf{E}_s+\mathbf{E}_{loc}$ contribute to the Hamiltonian in Eqn.~\ref{hfTerms} is \cite{marcusGndState}
\begin{equation}
    \frac{\hat{V}_{el}}{h} = d_\parallel E_{tot,z}\hat{S}_z^2-d_\perp E_{tot,x}(\hat{S}^2_x-\hat{S}^2_y) +d_\perp E_{tot,y}(\hat{S}_x\hat{S}_y+\hat{S}_y\hat{S}_x)
    \label{stark}
\end{equation}
Here, $d_\parallel$ and $d_\perp$ are coupling constants related to the NV electric dipole moment; $d_\parallel\ll d_\perp$ with $d_\parallel = 3.5 \times 10^{-3}$ Hz/(V/m) and $d_\perp = 0.17$ Hz/(V/m) \cite{eFieldSensing, VANOORT1990529}.  For the typical scale of sample electric fields, coupling to the NVs is small compared to sample magnetic fields of interest.  Hence electric fields do not cause noticeable shifts in ODMR resonances for most QDM magnetic imaging experiments. External temperature variations, e.g. from the sample, couple to the NV by the temperature dependence, $D(T)$, of the ZFS \cite{victor_Tdepend}. Changes in temperature of the diamond due to the sample temperature field will therefore result in a common mode shift of the ODMR resonance, which is distinct from the effect of magnetic fields. 

\subsection{NV Ground-State Hamiltonian}
Detecting the resultant spatial distribution of changes in the ODMR spectra across an NV ensemble due to spatially varying sample fields is the principle underlying QDM high-resolution imaging. The ground state Hamiltonian necessary to capture the relevant dynamics of single NVs for QDM imaging can be summarized by combining Eqns.~\ref{hfTerms}, \ref{strain}, \ref{Bsample}, and \ref{stark} 
\begin{align}
  \begin{aligned}
    \frac{\hat{H}_{gs}}{h} =\; & \frac{g_e\mu_B}{h}[(B_{0x}+B_{sx})\hat{S}_x+(B_{0y}+B_{sy})\hat{S}_y+(B_{0z}+B_{sz})\hat{S}_z] \quad & \text{Magnetic Terms} \\
   & + d_\parallel E_{tot,z}\hat{S}_z^2-d_\perp E_{tot,x}(\hat{S}^2_x-\hat{S}^2_y) +d_\perp E_{tot,y}(\hat{S}_x\hat{S}_y+\hat{S}_y\hat{S}_x) \quad & \text{Electric Terms} \\
   & +M_{z}\hat{S}_z^2+ M_{x}(\hat{S}^2_x-\hat{S}^2_y) + M_{y}(\hat{S}_x\hat{S}_y+\hat{S}_y\hat{S}_x) \quad & \text{Strain Terms} \\
   &+ D(T) \hat{S}_z^2  \quad & \text{ZFS, Temperature}
  \end{aligned}
  \label{Heffective}
\end{align}
Eqn.~\ref{Heffective} summarizes the interaction between an NV center and temperature, magnetic field, and electric field $\mathbf{E}_{tot}$. This equation demonstrates that NVs can in principle be used to image all of these quantities. For simplicity, Eqn.~\ref{Heffective} does not include the nuclear electric quadrupole interaction and the comparatively negligible terms of the crystal strain interaction. The hyperfine splitting terms from Eqn.~\ref{hfTerms}, also excluded for simplicity, are important and visibly evident in the hyperfine splitting of the ODMR resonances in Fig.~\ref{odmrExample}. Further simplifications can be made to Eqn.~\ref{Heffective} depending on the magnitude of the parameter of interest and the bias fields. 

\subsection{Behavior of NV Ensembles}
 QDMs use ensembles of NVs to obtain simultaneous wide-field of view measurements. Ensembles yield stronger signal than single NVs due to the larger number of NVs contributing photons to overall fluorescence, but introduce ensemble behavior that can worsen contrast compared to single NV performance. Intuition about single NVs also does not simply extend to the ensemble case: the NV spin-ensemble behavior can be substantively different from single NV behavior. The complex spin bath environment of the diamond contributes several mechanisms to NV spin ensemble dephasing and decoherence that ultimately limit the magnetic field sensitivity of NV ensembles. These ensemble dynamics are characterized by the longitudinal (spin-lattice) relaxation time $T_1$, the transverse (decoherence) relaxation time $T_2$, and inhomogeneously broadened transverse (dephasing) relaxation time $T_2^*$ \cite{slichter1996principles}. In particular, understanding and minimizing  ensemble NV dephasing is critical for high-performance broadband and static field magnetic imaging with QDMs, informing both diamond material design and quantum control techniques. This topic has been treated extensively in the literature \cite{Barry_SensitivityReport,P1DQ}, and will be discussed later in this paper, after introduction of NV measurement protocols.

\section{NV Measurement Protocols Relevant to QDMs}\label{protocols}
\begin{figure*}[ht]
\begin{center}
\begin{overpic}[width=0.9\textwidth]{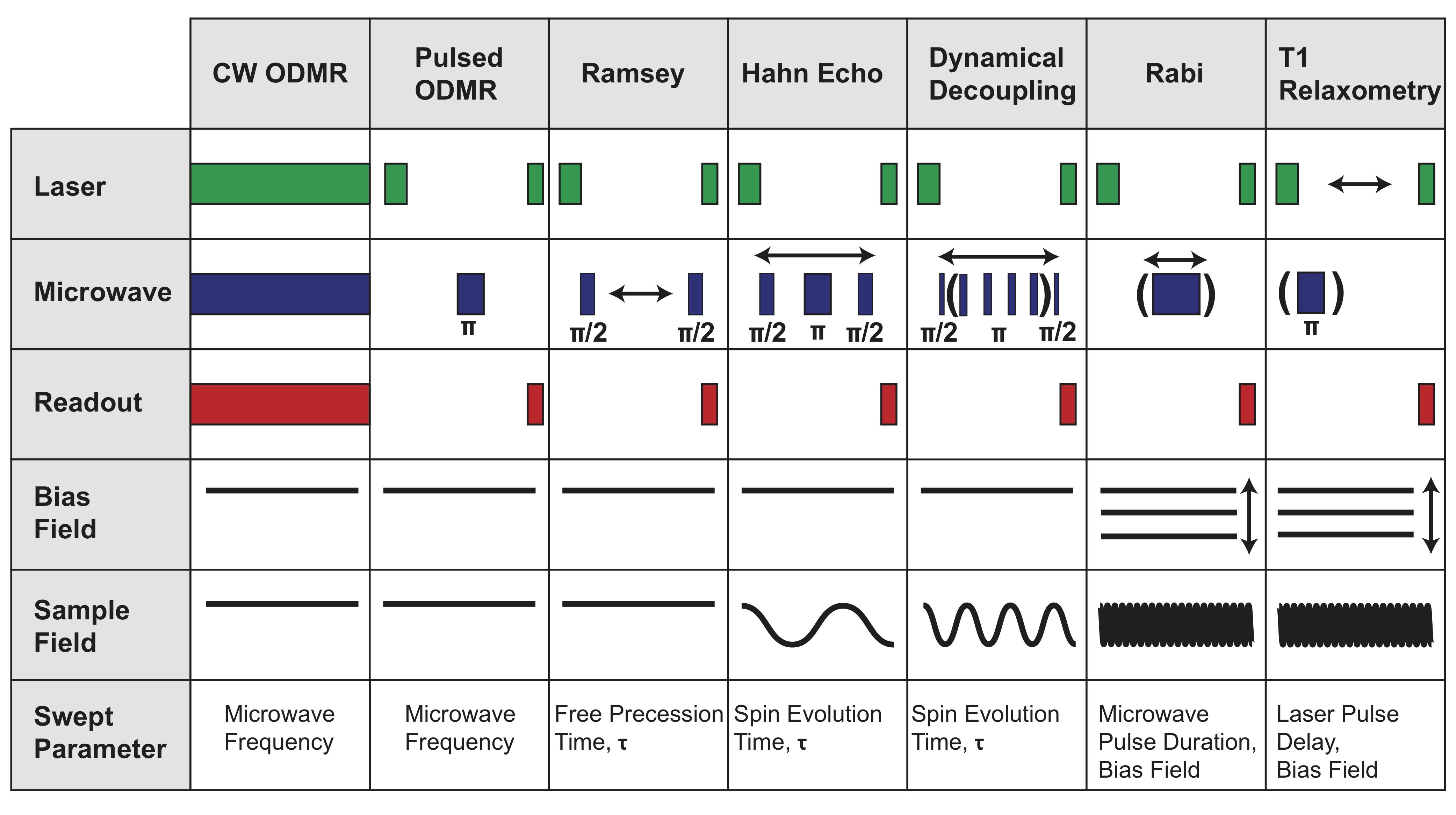}
\end{overpic}
\end{center}
\caption{\label{Techniques} NV Measurement Protocols.
 Schematic of timing and duration of laser pulses, MW pulses, and readout sequences relative to the field being sensed for common NV diamond protocols. Swept parameters are indicated by arrows. Straight lines for the bias and sample fields indicate static magnetic fields, including the swept static bias field for Rabi and $T_1$ Relaxometry; sinusoidal curves represent time dependent sample fields, which are very high frequency for Rabi and $T_1$ Relaxometry.}
\end{figure*}

The large toolbox of QDM sensing protocols allows for imaging magnetic fields over a wide range of characteristic timescales. Fig.~\ref{Techniques} shows a schematic of the most commonly used sensing protocols. The interplay and timing of the laser pulses and MW pulses determines the basic properties of the techniques introduced in this section.

\subsection{DC Magnetometry: Static and Low-Frequency Fields}
Three established QDM sensing protocols exist for measuring static (DC) and slowly-varying magnetic fields: continuous-wave (CW) ODMR, pulsed ODMR, and Ramsey magnetometry. These protocols have been used to sample time-varying magnetic fields up to 1 MHz in a single-pixel experiment \cite{shin:124519}.

\begin{figure*}[ht]
\begin{center}
\begin{overpic}[width=0.9\textwidth]{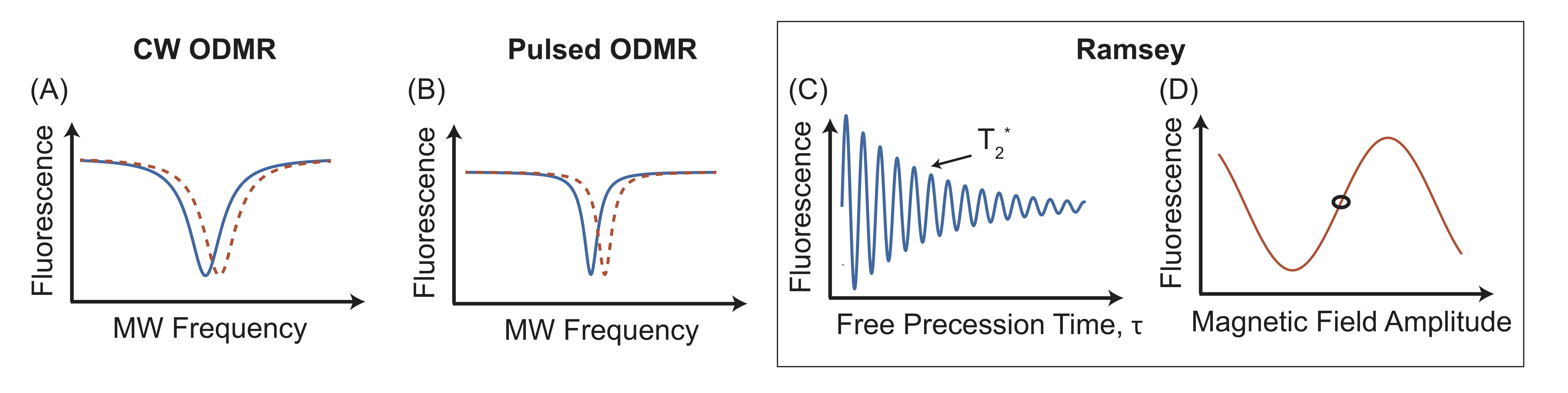}
\end{overpic}
\end{center}
\caption{\label{DCCartoon} DC Magnetometry Protocols.
 (A) Example CW ODMR lineshape before (blue) and after (red) change in magnetic field. (B) Example pulsed ODMR lineshape before (blue) and after (red) change in magnetic field. (C) Schematic Ramsey Free Induction Decay (FID) to determine dephasing time $(T_2^*)$ and optimal sensing time, $\tau_{sense}$. (D) Schematic Ramsey Magnetometry Curve. Free precession time is fixed to be the point of maximum slope of the FID curve closest to $T_2^*$, indicated by a black circle. Accumulated phase from sample field results in an oscillatory response of fluorescence with changing amplitude.}
 \end{figure*}

\subsubsection{CW ODMR}
CW ODMR is a robust and simple method that can image the vector components of a magnetic field in the QDM modality. Due to easy implementation, CW ODMR is the most common technique used for QDM applications. Continuous laser pumping, MW driving, and fluorescence readout are performed simultaneously, as shown in Fig.~\ref{Techniques}. The laser is used to both pump the NVs into the $ m_s= 0$ spin state and to probe spin states of the population via the NV fluorescence. The frequency of the MW drive is swept in time and synchronized with the readout. A decrease in fluorescence occurs when the MW frequency matches the NV resonance due to the spin state dependence of NV photon emission described in Section \ref{NVStates}.

Fig.~\ref{DCCartoon}(A) shows an example where a change in $\mathbf{B}_0$ shifts the line center of the resonance feature. For an NV ensemble, the resonance lineshape -- often modeled as a Lorentzian or Gaussian -- is parametrized by the center frequency, linewidth, and fluorescence contrast. The center frequencies of every NV resonance feature are fit to the appropriate Hamiltonian to extract the desired magnetic field, strain, temperature, and electric field. In a magnetic imaging experiment, this analysis yields $\mathbf{B}_0 + \mathbf{B}_s$, from which the magnetic field of the sample can be determined \cite{glennCancer,nv_bacteria,shin:124519}. 

Measuring the entire resonance spectrum in CW ODMR limits the sensitivity and the temporal resolution of the measurements, due to the significant fraction of experiment time spent interrogating with probe frequencies that yield no signal contrast. Sparse sampling of the resonance spectrum can improve the sensitivity of the measurement by minimizing dead time. An extreme version of sparse sampling can be achieved using a lock-in modality where the probe frequency is modulated between the points of maximum slope of an ODMR resonance feature \cite{barryNeurons}. This technique has been extended to monitor multiple ODMR features simultaneously to extract the vector magnetic field by modulating at different frequencies \cite{1803.03718}. Frequency-modulated ODMR has been performed with bandwidths up to 2 MHz, but was demonstrated on
a small volume and required high laser and MW intensity beyond that typically employed
with QDMs \cite{Barry_SensitivityReport}.

\subsubsection{Pulsed ODMR}
CW ODMR suffers from laser repumping of the NV spins through the entire measurement.
This simultaneous laser pumping and MW drive spoils the measurement sensitivity as a result of the competing processes of initializing the spin state (laser) and driving transitions
(MW drive) \cite{dreau_powBroad}. In order to mitigate this power broadening,
a pulsed ODMR protocol uses a temporally separated laser initialization, a MW control $\pi$ pulse, and a laser readout pulse as demonstrated in Fig.~\ref{Techniques}. This leads to the decreased linewidths shown in Fig.~\ref{DCCartoon}(B) as compared to CW ODMR. Alteration of the MW power changes the necessary duration of a $\pi$ pulse, and must be optimized to balance linewidth and contrast of ODMR resonance features \cite{dreau_powBroad}.

\subsubsection{Ramsey Spectroscopy}
Ramsey spectroscopy \cite{Ramsey} determines the magnitude of a DC magnetic field by measuring the relative phase accumulation between the different electronic spin states prepared in a superposition with a $\pi/2$ pulse  \cite{longT2, dima_magReview}. A green laser pulse initializes the spin state into the $m_s=0$ state to begin the sequence. Next, a resonant MW $\pi/2$ pulse prepares the spin into a superposition of the $m_s=0$ and $m_s=+1$ spin states (or $m_s=-1$ depending on the drive frequency). The system is allowed to evolve under the relevant Hamiltonian for a free precession time, $\tau.$ In the simplified scenario where the dynamics are dominated by the magnetic field, the NV superposition state accumulates a phase $\phi=2\pi\gamma (|\mathbf{B}_0+\mathbf{B}_s|)\tau$. A second MW $\pi/2$ pulse is applied to project the accumulated phase information onto the relative population of $m_s=0$ and $m_s=+1$ spin states. A second laser pulse is applied to measure the spin state population through the spin dependent fluorescence of the NV. 

To obtain information about the magnetic field, $|\mathbf{B}_0+\mathbf{B}_s|$, a Ramsey pulse sequence is repeated several times, sweeping the free precession interval such that each measurement is taken for different $\tau$. The resultant fluorescence contrast signal as a function of $\tau$ is known as the Ramsey fringes, illustrated in Fig.~\ref{DCCartoon}(C). By taking a Fourier transform of these fringes, one can locate the position of the dominant frequencies and determine deviations from those set by the bias magnetic field that result from the sample field. 

Mapping out the Ramsey fringes is inefficient with respect to speed of the measurement, similar to the inefficiency of the frequency sweep for ODMR. Instead of mapping out the full fringe and taking the Fourier transform, the free precession time, $\tau_{sense}$, is fixed to sample the Ramsey fringe at the point of maximum sensitivity, the point of maximum slope closest to $T_2^*$. This process maps out a magnetometry curve illustrated in Fig.~\ref{DCCartoon}(D). The steeper the slope of the magnetometry curve, the more sensitive the protocol. 

A key feature of Ramsey magnetometry is having both the laser and MWs switched off when the NV electronic spin is accumulating phase via interaction with the magnetic field. The Ramsey protocol is consequently not vulnerable to the power broadening that impacts CW ODMR, and allows use of high MW and laser power to increase the sensitivity \cite{Barry_SensitivityReport}. 
Other benefits of Ramsey magnetometry over CW and pulsed ODMR is that it more efficiently leverages protocols that mitigate dephasing such as spin bath driving and is compatible with sensing in the double quantum basis \cite{P1DQ}. 

\subsection{AC Magnetometry: Narrowband Fields}
A QDM can measure narrowband oscillating magnetic fields using AC magnetometry sequences, including Hahn Echo and Dynamical Decoupling. These pulse sequence protocols act as frequency filters and allow the QDM to operate as sensitive lock-in detector \cite{dima_magReview} of these AC fields. The frequency range of narrowband signals that are detectable with NV AC magnetometry is $\sim1$ kHz to $\sim10$ MHz, limited at the low end by NV decoherence and at the high-end by the amplitude of fast MW pulses that can be realistically applied to an NV ensemble.

\begin{figure*}[ht]
\begin{center}
\begin{overpic}[width=0.7\textwidth]{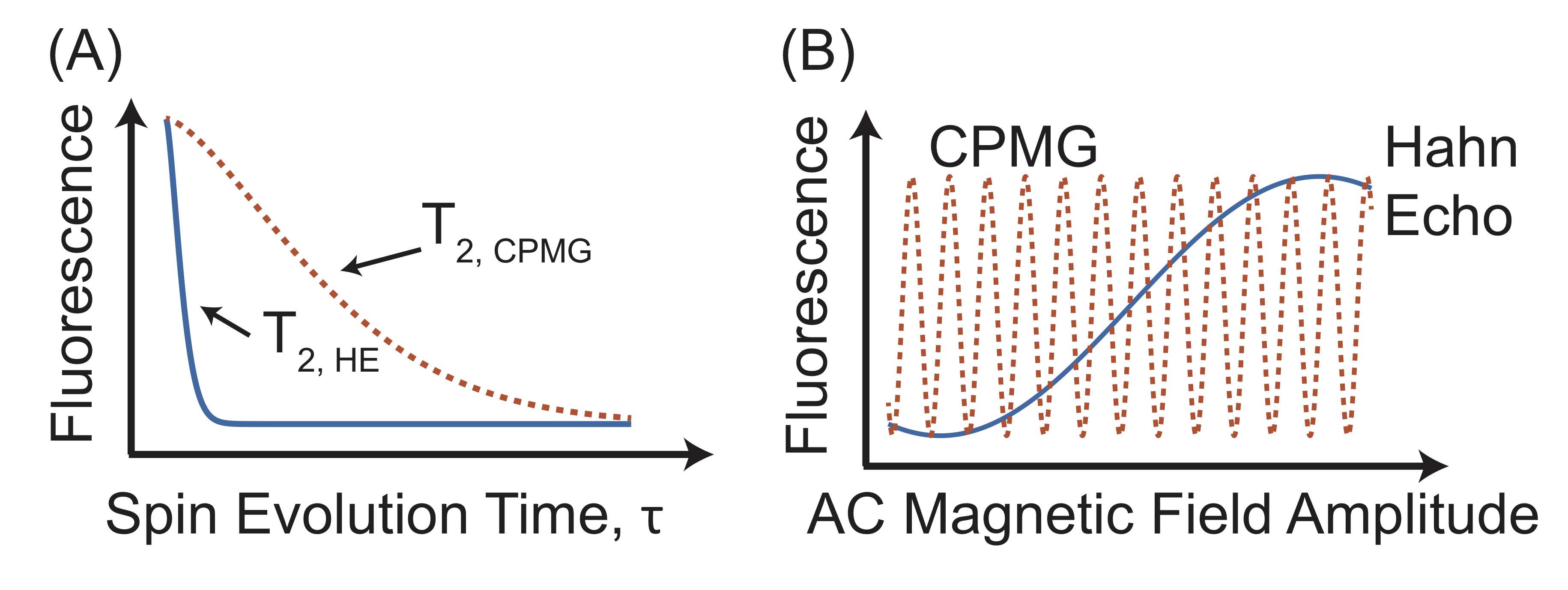}
\end{overpic}
\end{center}
\caption{\label{ACCartoon} AC Magnetometry Protocols.
 (A) Schematic $T_2$ decoherence curves for Hahn Echo and CPMG-32 sequences. Improved decoupling from interactions with the spin bath environment results in an extended CPMG $T_2$ decoherence time compared to the Hahn Echo $T_2$ (B) Schematic magnetometry curves. Longer $T_2$ for CPMG leads to increased magnetic field sensitivity, as indicated by the slope of the CPMG magnetometry curve. Hahn Echo is much less sensitive,  exhibiting a similar oscillation to CPMG over a much larger magnetic field amplitude range.}
\end{figure*}

\subsubsection{Hahn Echo}
The addition of a $\pi$ pulse into the middle of a Ramsey sequence mitigates environmental perturbations that are slow compared to the free precession interval between pulses \cite{dima_magReview}. This pulse sequence is known as the Hahn Echo sequence \cite{longT2, deLange2010}, and results in the refocusing of NV ensemble dephasing such that the limiting measurement timescale becomes the decoherence time $T_2$ rather than the dephasing time $T_2^*$.  The consequence is improved magnetic field sensitivity (discussed in Section 4.1), especially for lower frequency signals, because $T_2$ typically exceeds $T_2^*$ by over an order of magnitude \cite{2019arXiv190408763B}. Fig.~\ref{ACCartoon}(A) demonstrates a decoherence curve when using a Hahn Echo pulse protocol.  The spacing between MW pulses acts as a narrowband filter in frequency space. The width of this filter is given by the filter response function \cite{Sarma}. Hahn Echo uses only one $\pi$ pulse and therefore has a fairly broad filter allowing for sensing of a wide bandwidth of magnetic field frequencies.  

To optimally sense external oscillating fields, the spin evolution time is set to be $\sim T_2$; however, the frequency of the sensed magnetic field can lead to operation with a non-optimal $\tau$. \cite{dima_magReview} For a fixed spin evolution time, a change in magnetic field will lead to a difference in phase accumulation that maps onto the total fluorescence, Fig.~\ref{ACCartoon}(B). 

\subsubsection{Dynamical Decoupling}

Building upon the Hahn Echo sequence, dynamical decoupling techniques commonly apply multiple refocusing pulses with spacing determined by the period of the sample field. \cite{dima_magReview, LinhDD} These additional refocusing pulses result in an advantageous extension of the decoherence time by narrowing the width of the filter response function and reducing sensitivity to magnetic noise outside the bandwidth. In particular, decoupling of the NV from spin-bath-induced magnetic noise improves with additional pulses at the trade-off of making the technique sensitive to a narrower range of frequencies \cite{Sarma, deSousaBook}. The extension in the decoherence time, Fig.~\ref{ACCartoon}(A), can lead to a dramatic improvement in magnetic field sensitivity Fig.~\ref{ACCartoon}(B). Dynamical decoupling also increases the time during which NVs can interrogate the sample field $B_s$, optimally towards the extended decoherence time.

The Carr-Purcell-Meiboom-Gill (CPMG) pulse sequence is a dynamical decoupling sequence employing $\pi$ pulses which rotate the NV about the same axis as it is polarized by the initial $\pi/2$ pulse. Another common sequence, XY8, extends this by choosing the rotation axis for each $\pi$ pulse in order to suppress the effects of pulse errors. A large family of similar sequences exist, many well known in NMR, to improve NV sensing through more efficient robust control of the NV electronic spin state \cite{1705.02262}.

\subsection{Resonant Coupling to External GHz Fields}
\begin{figure*}[ht]
\begin{center}
\begin{overpic}[width=0.8\textwidth]{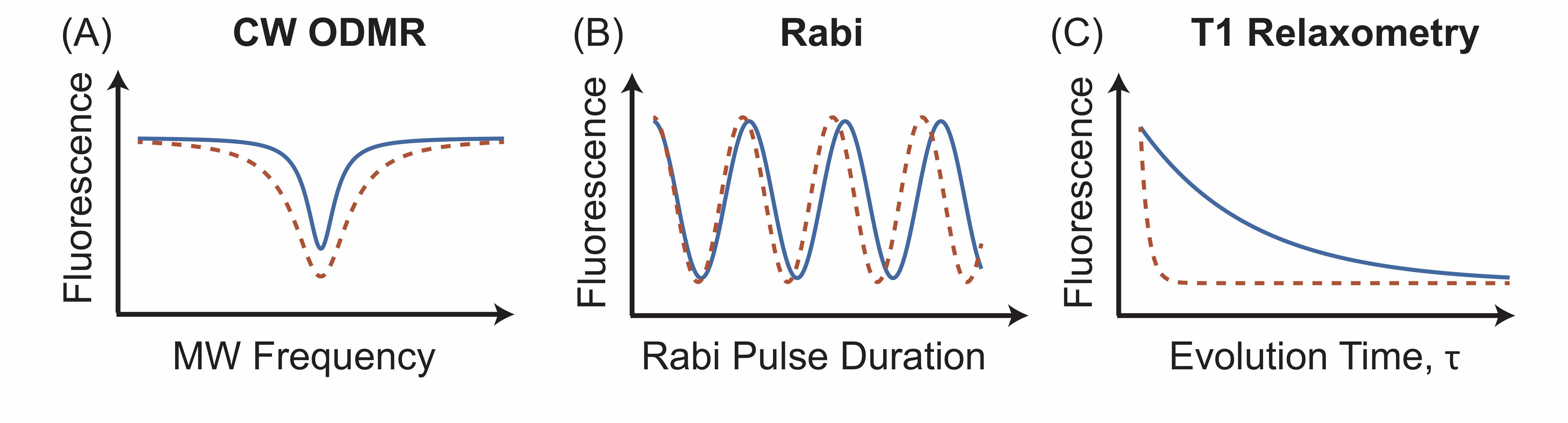}
\end{overpic}
\end{center}
\caption{\label{GHzCartoon} GHz Magnetometry Protocols.
 (A) CW ODMR broadening: increased MW power will increase ODMR fluorescence contrast and linewidth (red). (B) Rabi Oscillation: increased amplitude of the MW field will increase the Rabi frequency (red). (C) $T_1$ relaxometry: phonon limited $T_1$ decay rate (blue) is increased (red) by high frequency magnetic noise near the NV resonance frequency.}
\end{figure*}

Applications that require measurement of GHz scale oscillations can leverage interactions between the NV and magnetic signals near the NV resonance as a probe \cite{linboMWimging, maletinskyMWimger,T1flowimaging}. CW ODMR constitutes a simple protocol that can be used in this manner. Measurements of the contrast and linewidth enable the determination of the optical power and MW power that power broaden the lines, Fig.~\ref{GHzCartoon}(A), in addition to other mechanisms that contribute to the ensemble inhomogeneous dephasing \cite{dreau_powBroad}.  However, this method is not very sensitive and difficult to quantify due to the various ways the contrast and linewidth can vary over a field of view \cite{QDM1ggg}. Alternative methods to CW ODMR include Rabi Driving and $T_1$ Relaxometry.

\subsubsection{Rabi Driving}
Use of a Rabi sequence provides a more direct way to determine local magnetic fields oscillating at or near GHz frequencies as compared to CW ODMR \cite{maletinskyMWimger}. Similar to previously discussed protocols, the NV spin state is initiated to the $m_s=0$ state  with a green laser light. A MW drive is left on for a varying amount of time. If the MW drive is on resonance with the NV spin state transitions, for example between $m_s=0$ and $m_s=+1$ states, the population will be driven back and forth between the spin states. The strength of this GHz drive determines the rate at which the transition is driven. This rate is called the Rabi frequency, and scales with the square root of the input microwave power. Fig.~\ref{GHzCartoon}(B) illustrates the increase in Rabi frequency as a function of increasing amplitude of the MW driving field.

\subsubsection{$T_1$ Relaxometry}
When the NV is initially polarized into the $m_s=0$ state with green illumination, there is a characteristic timescale over which the spin population decays back to a thermally mixed state. This timescale is the longitudinal (spin-lattice) relaxation time $T_1$ and can be up to 6 ms when dominated by phonon interactions at room temperature \cite{hollenberg_DD}. However, $T_1$ can be spoiled by the presence of magnetic frequency noise or other paramagnetic spins at the NV resonance frequency, as shown in Fig.~\ref{GHzCartoon}(C) \cite{imaging_mag}. The local bias field can be swept to change the frequency of noise the measurement to which it is sensitive. 

\section{QDM Performance}
QDM performance characteristics include magnetic field sensitivity, temporal resolution, frequency bandwidth, spatial resolution, and field of view of the sample field. These characteristics depend on the sensing protocol of the QDM discussed in Section \ref{protocols}, which in turn are determined by the spectral and spatial qualities of the sample fields to be imaged. This section focuses on the physical limits to performance; the performance impact resulting from use of different experimental components for the QDM is treated in Section \ref{components}.

\subsection{Magnetic Field Sensitivity} 
The minimum detectable field difference is defined as the change in magnetic field magnitude, $\delta B$ for which the resulting change in a given measurement of the field equals the standard deviation of a series of identical measurements. However, characterizing the minimum detectable field difference must consider the total measurement duration, as well as the total number of NVs that contribute to the measurement for meaningful determination of sensor performance. The magnetic field sensitivity scales as the square root of the number of detected photons. The number of photons collected over a unit time from a unit volume of NVs increases proportionally with time and volume. To account for measurement time, sensitivity is represented as $\eta = \delta B \sqrt{t_{meas}}$ with units of T Hz$^{-1/2}$, where $t_{meas}$ is the total measurement time.  To account for the number of NV spins required to reach a given sensitivity, a volume-normalized sensitivity is defined as $\eta_{vol} = \eta \sqrt{V}$ with units T $\upmu$m$^{3/2}$ Hz$^{-1/2}$, where $V$ is volume and for a fixed density of NVs \cite{RomalisSERF,OptMag}. 

\begin{table}[]
\resizebox{\textwidth}{!}{%
\begin{tabular}{|l|l|l|l|l|l|l|l|l|l|}
\hline
Diamond [N] & $n_N$  & NV/N &  $n_{NV, SA}$ & \begin{tabular}[c]{@{}l@{}}Single Axis\\ Counts/s\end{tabular} & \begin{tabular}[c]{@{}l@{}}Photon Rate\\ (Counts/s)\end{tabular} & \begin{tabular}[c]{@{}l@{}}Counts\\Per Readout\end{tabular} & Contrast & $T_2^*$ ($\upmu$s)  & $T_2$ ($\upmu$s) \\ \hline
1 ppm   & 1.76*10\textasciicircum{}5 & 0.1    & 4.4*10\textasciicircum{}3 & 10\textasciicircum{}5        & 4.4*10\textasciicircum{}8  & 132     & 5\%     & 10.0 & 160 \\ \hline
20 ppm   & 3.52*10\textasciicircum{}6 & 0.1    & 8.8*10\textasciicircum{}4 & 10\textasciicircum{}5      & 8.8*10\textasciicircum{}9    & 2640  & 5\%      & 0.50 & 8.0 \\ \hline
\end{tabular}}

\caption{\label{Diamonds} Simulated diamond properties. Properties of two notional diamonds used for performance simulations for a 1 $\upmu \text{m}^3$ QDM voxel. [N] is the concentration of nitrogen in the diamond. $n_N$ is the number of nitrogen atoms per 1 $\upmu \text{m}^3$ voxel, NV/N = 10$\%$ of the N atoms are NV centers. A single NV axis is considered, giving $n_{NV,SA}$ with 10$^5$ fluorescence counts/s for each NV in a 1 $\upmu \text{m}^3$ QDM voxel. More NV centers per voxel increases magnetic field sensitivity because the rate of photons emitted scales with $n_{NV^-}$. Counts per readout are for an assumed readout time of 300ns. Assumed scaling of $T_2^*$ and $T_2$ are from Ref \cite{2019arXiv190408763B}. }
\end{table}

\begin{figure*}[ht]
\begin{center}
\begin{overpic}[width=0.75\textwidth]{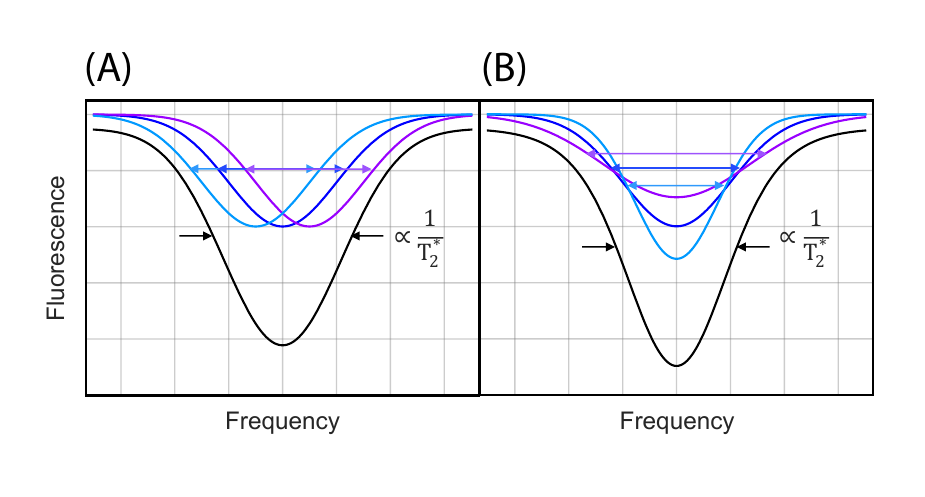}
\end{overpic}
\end{center}
\caption{\label{Broadening}
Mechanisms for Ensemble Broadening. Different colors represent resonance features from different NV sub-ensembles that contribute to the measured resonance linewidth from the entire NV ensemble. (A) Differences in linewidth center frequency due to variations in local environment, e.g., magnetic field and/or strain gradients, and (B) differences in contrast and linewidth due to dephasing from variations in T$_2^*$, and/or power broadening. From Ref. \cite{Barry_SensitivityReport}.}
\end{figure*}

CW ODMR magnetometry is the most widely used QDM measurement technique due to its simplicity. The sensitivity of a CW ODMR magnetometry sequence is characterized by the slope of the resonance line, $\partial I/\partial \nu_0$ with fluorescence intensity $I$ and frequency $\nu_0$, and the rate of photon detection from a cubic micron of NVs, $R$. The CW ODMR, shot-noise-limited sensitivity is 
\begin{equation}
    \eta_{\text{CW ODMR}} \approx 2\pi\frac{\hbar}{g_e\mu_B}\frac{\sqrt{R}}{\text{max}|\partial I/\partial \nu_0|} = \frac{8 \pi} {3 \sqrt{3}} \frac{\hbar}{g_e \mu_{B}}\frac{\Delta \nu}{C \sqrt{R}}
    \label{SensivityODMR}
\end{equation}
where $C$ is the contrast and $\Delta \nu$ is the linewidth of the ODMR resonance. The resonance line shape is typically fit by a Lorentzian, giving the $4/(3\sqrt{3})$ factor for the maximum slope. The relationship between the ODMR linewidth and the previously defined dephasing time $T_2^*$ is approximated by $T_2^*=(\pi \Delta\nu)^{-1}$ \cite{P1DQ, Barry_SensitivityReport,PhysRevLett.102.237601, 2019arXiv190408763B}. When performing ensemble measurements, many mechanisms can contribute to the linewidth as demonstrated in Fig.~\ref{Broadening}.

\begin{figure*}[ht]
\begin{center}
\begin{overpic}[width=0.95\textwidth]{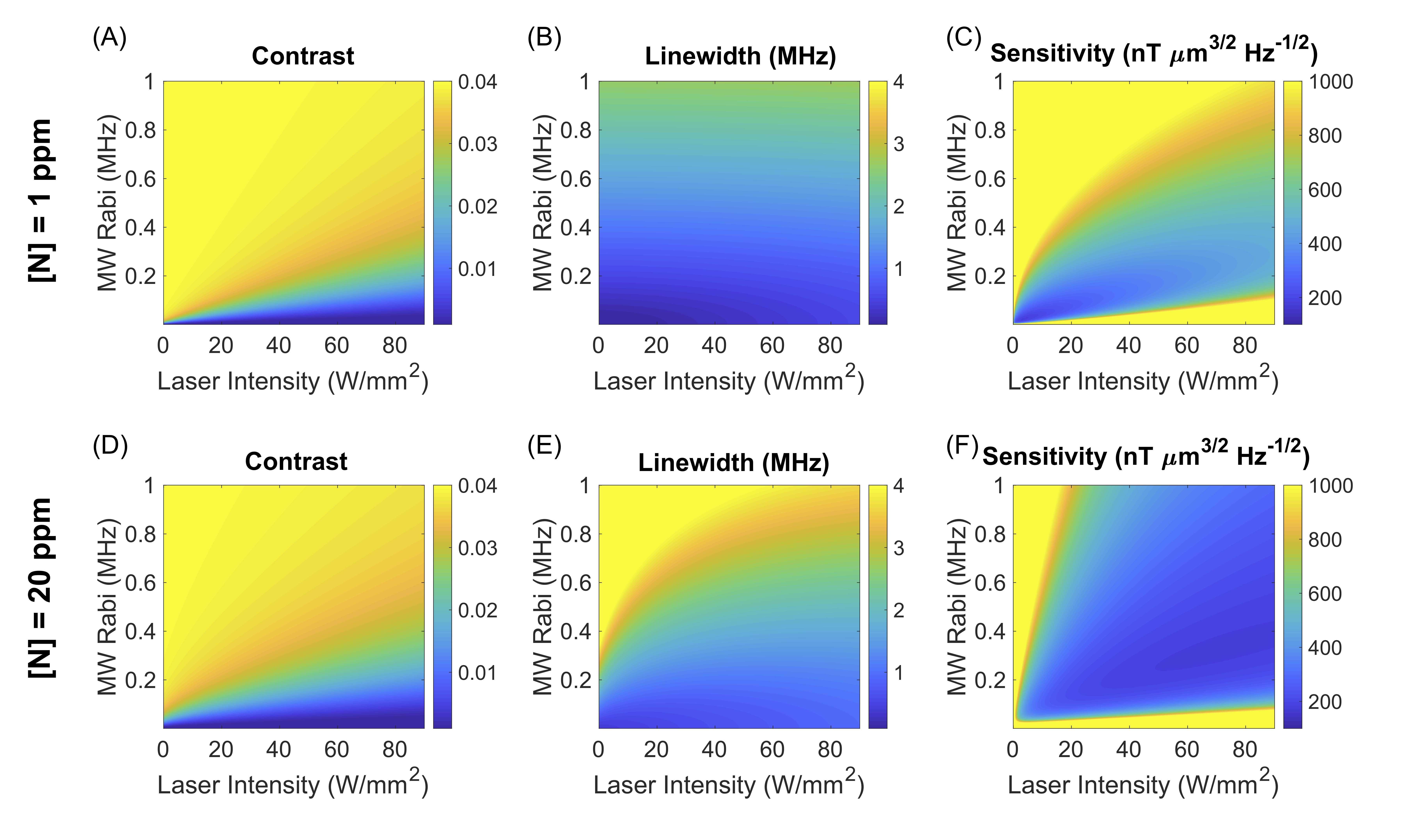}
\end{overpic}
\end{center}
\caption{\label{ODMRSens}
Ensemble CW ODMR Sensitivity Analysis. Simulations of CW ODMR contrast, linewidth, and volume-normalized magnetic field sensitivity $\eta_{vol}$ as a function of laser intensity and MW Rabi (which scales as the square root of the input MW power) with parameters from Table \ref{Diamonds} for a diamond with 1 ppm of nitrogen (Top Row), and a diamond with 20 ppm nitrogen (Bottom Row). Laser intensity scale assumes saturation intensity of 0.9 MW/$\text{cm}^2$.}
\end{figure*}

The sensitivity of CW ODMR magnetometry is limited by laser and MW ODMR lineshape power broadening. Solving the Bloch equation for a simplified two-level model yields  the contrast, linewidth, and volume-normalized magnetic sensitivity, shown in Fig.~\ref{ODMRSens}. The calculations are based on \cite{dreau_powBroad} for CW ODMR  using parameters from Table \ref{Diamonds}, Fig.~\ref{ODMRSens} displays a broad range of laser and MW powers to indicate how these affect the  sensitivity. The tradeoff between laser and MW power limits the achievable volume normalized-sensitivity of CW ODMR, precluding simultaneous optimal contrast and narrow linewidth. Applications that require higher temporal and spatial resolution must use techniques more sensitive than CW ODMR.

\begin{figure*}[ht]
\begin{center}
\begin{overpic}[width=0.95\textwidth]{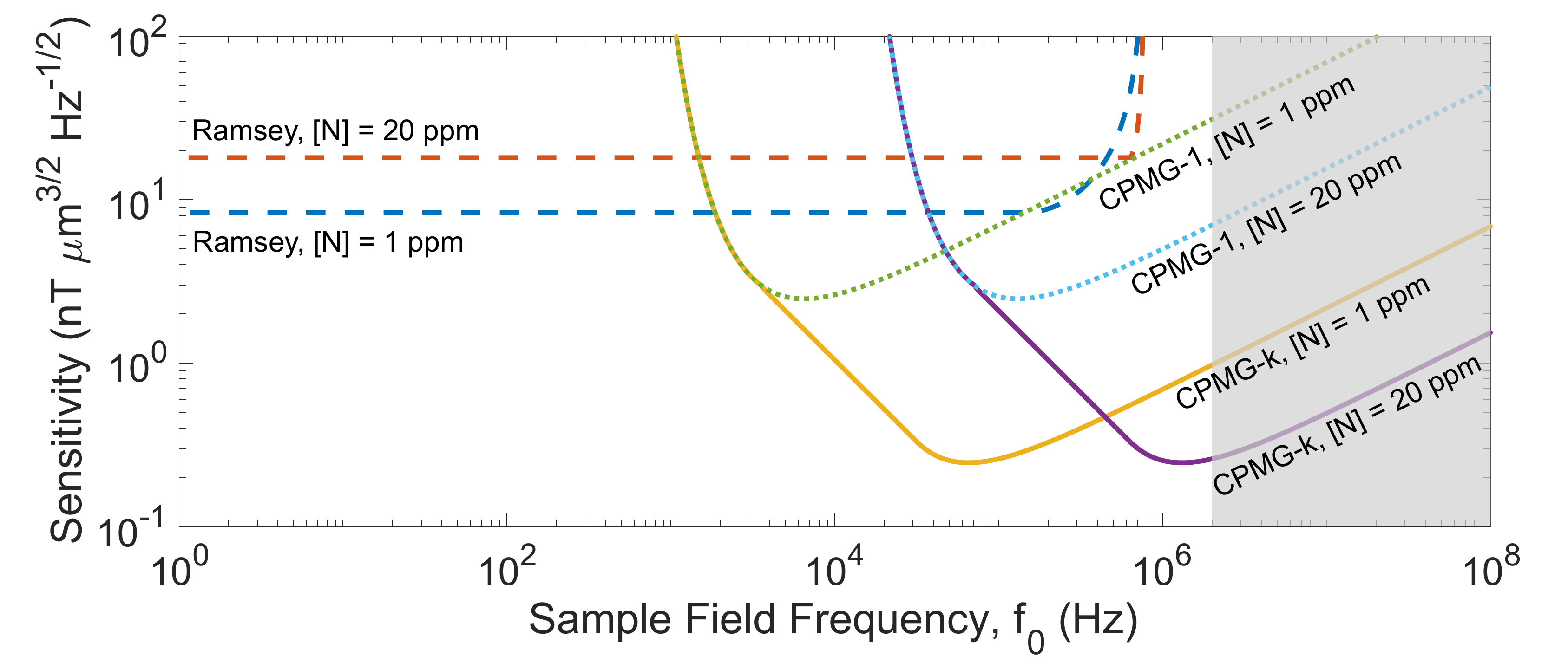}
\end{overpic}
\end{center}
\caption{\label{Sensitivity}
Frequency Dependence of QDM Volume-Normalized Sensitivity. Achievable volume-normalized magnetic field sensitivity as a function of the sample field frequency for DC broadband and AC narrowband QDM protocols. Calculations use parameters from Table \ref{Diamonds}. Ramsey is broadband, and is sensitive to magnetic fields of differing frequencies without requiring changes in the pulse sequence. CPMG is narrowband and requires a change in the pulse sequence based on the field frequency being sensed in order to maintain optimal sensitivity. The grey region indicates high sample frequencies where experimental requirements on MW pulses and power become technically challenging. Dotted lines are for a single pulse, which achieves the same sensitivity as a Hahn Echo sequence. Solid lines are for CPMG-k protocols limited up to 1024 pulses. }
\end{figure*}

Ramsey magnetometry achieves the best DC magnetic field sensitivity of the QDM protocols due to its pulse scheme: the NV spins interrogate the sample fields during an interval without simultaneous interaction of the optical and MW driving fields. The shot-noise-limited sensitivity for DC magnetic fields using a Ramsey pulse is \cite{Barry_SensitivityReport}
\begin{equation}
    \eta_{\text{Ramsey}} \approx \underbrace{\frac{\hbar}{g_e\mu_B}\left(\frac{1}{ \Delta m_s\sqrt{N\tau}}\right)}_{\text{Spin Projection Noise }}\underbrace{\left(\frac{1}{e ^{-(\tau/T_2^*)^p} }\right)}_{\text{Spin Dephasing}}\underbrace{\sqrt{1+\frac{1}{C^2n_{avg}}}}_{\text{Readout}} \underbrace{\sqrt{\frac{t_I+\tau+t_R}{\tau}}}_{\text{Overhead Time}}
    \label{SensivityRamsey}
\end{equation}
where $N$ is the number of non-interacting NVs contributing to the measurement $S=1/2$ spins, and $\Delta m_s$ is the generalization to greater-than-one spin state difference used for measurement (e.g., $\Delta m_s=2$ for the NV $m_s=-1$ to $m_s=1$ transition when operating with a double-quantum coherence \cite{P1DQ}), $C$ is the resonance contrast, $n$ is the average number of photons collected per NV per measurement, $\tau$ is the spin interrogation time, and $t_I$ and $t_R$ are the optical spin-state initialization and readout times respectively, ($t_{meas}=t_I+\tau+t_R $). The spin-projection-noise-limited sensitivity is given by the first two terms of Eqn.~\ref{SensivityRamsey}. It is evident that longer interrogation time, $\tau$ and larger number of spins, $N$ allow for better sensitivity to small magnetic fields. However, several factors cause Ramsey magnetometry to fall short of this limit: a decrease in sensitivity due to spin dephasing with characteristic time $T_2^*$ is accounted for in the exponential term with parameter $p$ depending on the origin of dephasing; imperfect readout contributes the first square root term; and the reduced fraction of total measurement time allocated for spin interrogation due to the overhead time from $t_I$ and $t_R$ is accounted for in the last term. Optimal DC sensitivity is achieved for $\tau \sim T_2^*$ \cite{Barry_SensitivityReport}. Fig.~\ref{Sensitivity} compares the sensitivity of Ramsey magnetometry as a function of the frequency of the field being measured for the two diamonds in Table \ref{Diamonds}.

The sensitivity for measurement of AC magnetic fields using the Hahn echo protocol is 
\begin{equation}
    \eta_{\text{Hahn Echo}} \approx \underbrace{\frac{\pi}{2}\frac{\hbar}{g_e\mu_B}\left(\frac{1}{ \Delta m_s\sqrt{N\tau}}\right)}_{\text{Spin Projection Noise }}\underbrace{\left(\frac{1}{e ^{-(\tau/T_2)^p} }\right)}_{\text{Spin Decoherence}}\underbrace{\sqrt{1+\frac{1}{C^2n_{avg}}}}_{\text{Readout}} \underbrace{\sqrt{\frac{t_I+\tau+t_R}{\tau}}}_{\text{Overhead Time}}
    \label{SensitivityEcho}
\end{equation}
Hahn echo magnetometry builds on the Ramsey protocol as discussed in Section 3.2.1, resulting in similar physics underlying the AC magnetic field sensitivity to that of DC fields. The additional MW $\pi$ pulse in the Hahn Echo sequence refocuses the dephasing NV ensemble such that the sensing duration, $\tau$ approaches the spin decoherence time $T_2$. Because $T_2$ is at least an order of magnitude longer than the spin dephasing time, $T_2^*$, the sensing duration increases translating to an improvement in sensitivity. AC sensing protocols are thus limited by $T_2$, whereas DC sensing protocols are limited by $T_2^*$; because $T_2\gg T_2^*$, the AC protocols can generally achieve better sensitivity than DC protocols. However, the benefit of being $T_2$ limited can be degraded by coherent interactions between the NV spin ensemble and other spin impurities that decrease the $T_2$ coherence time. The optimal spin interrogation time $\tau$ for Hahn echo magnetometry is $\tau\sim T_2$; additionally, $\tau$ should match the period of the AC magnetic field, $T_{AC}$. Consequently, maximal sensitivity is achieved for AC magnetic fields with $T_{AC}\sim T_2$ of the diamond.
  
CPMG pulse sequences improve the sensitivity by extending $T_2$ even further \cite{Barry_SensitivityReport}
\begin{equation}
    \eta_{\text{CPMG}}\approx \frac{\pi}{2} \frac{\hbar}{g_e\mu_B}\left(\frac{1}{ \Delta m_s\sqrt{N\tau}}\right)\left( \frac{1}{e^{-(k^{-s}\tau/T_2)^p}}\right) \sqrt{1+\frac{1}{C^2n_{avg}}}\sqrt{\frac{t_I+\tau+t_R}{\tau}}
    \label{cpmg}
\end{equation}
where $k$ is the number of pulses, and $\tau=k/(2f_0)$ is the full spin evolution time, and $f_0$ is the frequency of the sample field. The optimal number of pulses for a given frequency is given by $k_{opt}=\left( \frac{1}{2p(1-s)}(2T_2f_0)^p\right)^{1/(p(1-s))}$, with s $\sim 2/3$ and p $\sim 3/2$, and is set mostly by the spin bath dynamics \cite{LinhDD}. The measurement time increases linearly with increased number of pulses, whereas the decoherence time $T_2$ increases sublinearly; the optimal number of pulses balances these effects \cite{Barry_SensitivityReport}.  Extensions of Eqn.~\ref{cpmg} exist to take into account multi-pulse dynamical decoupling protocols. 

\subsection{Temporal Resolution and Frequency Bandwidth}  
QDM temporal resolution is defined as the time required between subsequent measurements of the sample field. The physical limitation determining the fastest temporal resolution is set by the time it takes for the NVs to react to a change in the sample field. The temporal resolution can never be faster than $\sim$5 MHz (the maximum optical pumping rate), which is limited by the $^1E$ metastable state lifetime. The same is true for pulsed measurements, since NVs are optically reinitialized to the $m_s=0$ state before each measurement. For a measurement with continuous laser illumination and MW field, the NV temporal resolution is set by the optical pumping rate and the MW Rabi frequency. There is also a practical limit to the temporal resolution, set by signal-to-noise ratio (SNR) tolerance: faster temporal resolution gives worse SNR per measurement. 

The NV sensor frequency bandwidth is the range of sample frequencies that can be interrogated with the same experimental protocol. A DC magnetometry experiment has a frequency resolution spanning DC to the NV temporal bandwidth cutoff. A dynamical decoupling AC magnetometry experiment has an approximate frequency bandwidth that is roughly equal to $1/T_{tot}$ (the Fourier limit), where $T_{tot}$ is the duration of the dynamical decoupling pulse sequence. An AC magnetometry measurement based on driving the spin population between $m_s$ sublevels (Rabi) or spoiling of the initialized spin state ($T_1$ relaxometry) has a frequency bandwidth corresponding to the NV resonance linewidth; i.e, the frequency span over which the NVs are on resonance with the MW field, which is $>1/(\pi T_2^*)$.

\begin{figure*}[ht]
\begin{center}
\begin{overpic}[width=0.95\textwidth]{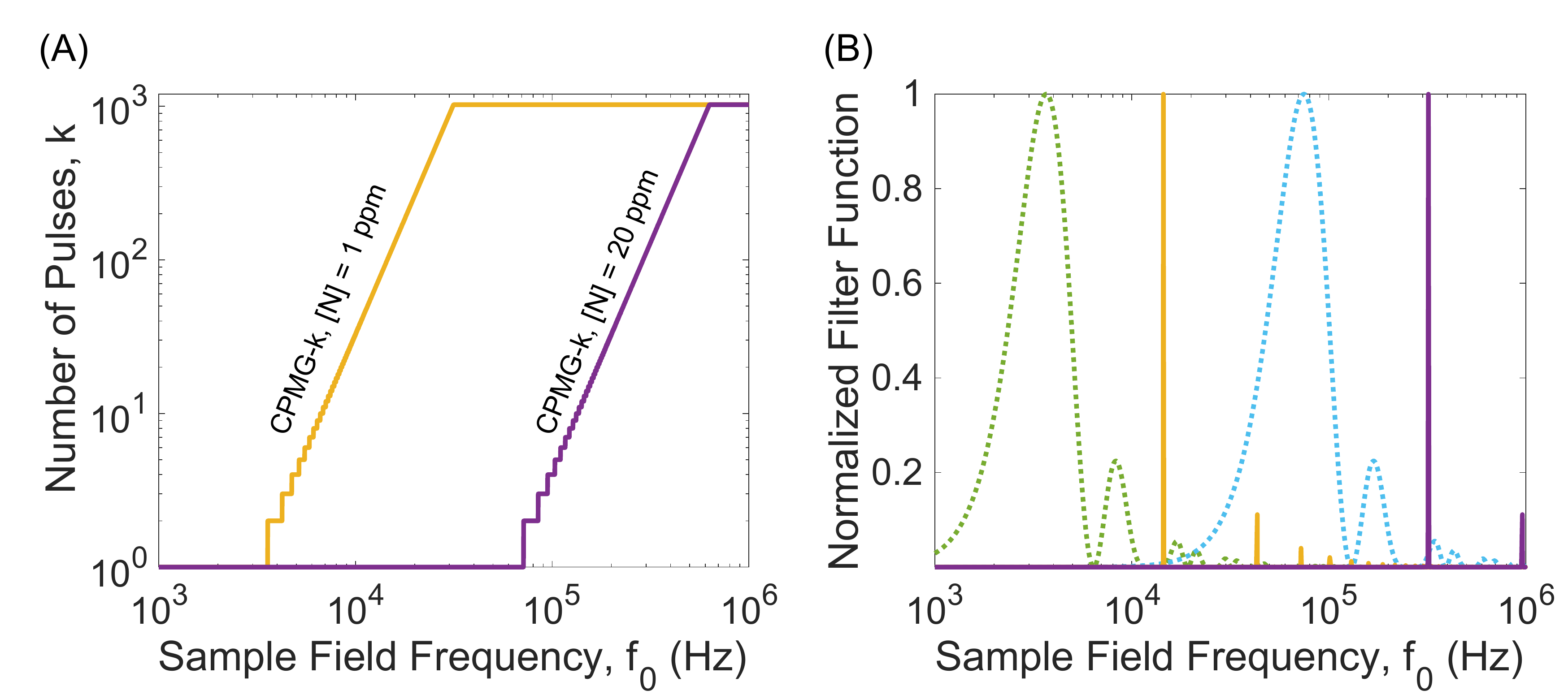}
\end{overpic}
\end{center}
\caption{\label{BandWidth}
CPMG Protocol Bandwidth. (A) The optimal number of pulses for the CPMG protocol changes with the sample field frequency. (B) Example filter functions $S(f)$, at the most sensitive sample frequencies for each of the CPMG curves in Fig.~\ref{Sensitivity}. Dotted lines represent the response for a 1 pulse CPMG. Solid lines are for the most sensitive center sample field frequencies for CPMG limited to 1024 pulses. The solid lines are $\sim$1000 times narrower than the dotted lines due to having $\sim$1000 times more pulses.}
\end{figure*}

AC magnetometry sequences that are based on pulse control of the NV spin state have a frequency bandwidth dictated by the filter function, $S(f)$, for the specific pulse sequence being used. The center frequency and bandwidth are defined by the number of pulses, $k$, and the spacing between $\pi$ pulses, $\tau$ \cite{Sarma}. The center frequency of the filter is given by $f_0= 1/2\tau$ where $\tau$ is the spacing between $\pi$ pulses. For a sequence of $k$ pulses, with total measurement time $T=k\tau$, the width of the filter function is given by $\Delta f =1/T =  1/k\tau$. The filter function $S(f)$ depends upon a protocol-specific response $F(fT)$

\begin{align}
    S(f)=2F(f T)/(2\pi f)^2,
    \label{filter}
\end{align}
\noindent
where an example response function for the CPMG protocol is \cite{Sarma}

\begin{align}
    & F_{\text{CPMG-k}}(fT)=8\sin^4\left(\frac{2\pi f T}{4k}\right) \sin^2\left(\frac{2\pi f T}{2}\right) / \cos^2\left(\frac{2\pi f T}{2k}\right).
    \label{filterCPMG}
\end{align}

Fig.~\ref{BandWidth}(A) demonstrates the need to change in number of pulses in order to operate at the sensitivity limit shown in the CPMG curves in Fig.~\ref{Sensitivity}. Fig.~\ref{BandWidth}(B) gives filter functions for the most sensitive points along the curves presented in Fig.~\ref{Sensitivity}.

It is tempting to conflate temporal bandwidth and frequency bandwidth, but they in fact represent different characteristics. For example, an NV $T_1$ measurement can detect magnetic noise across a few MHz frequency bandwidth around a central frequency ranging from near zero to many GHz (depending on the applied $\mathbf{B}_0$), but measurements may only be repeatable at $<1$ kHz (temporal resolution). Only in the case of DC magnetometry protocols do the temporal and frequency bandwidth correspond to the same sensor property.

\subsection{Spatial Resolution and Field of View} 
QDM magnetic imaging seeks to resolve magnetic fields with high spatial variation over a wide-field of view, and to successfully invert the magnetic field measurements to generate a map of closely-separated magnetic sources. Both goals have fundamental and sensor-specific limitations. It is ideal to operate at the limit of magnetic field inversion, and not to be limited by the sensor properties such as resolution and field of view.

The magnetic inversion problem does not generally have a unique solution. Only if the current distribution is limited to two dimensions (2D) can the inverse problem be solved uniquely from a planar measurement of the magnetic field. A magnetometer must sample the field at discrete points in a 2D plane with a sufficient sampling density to recover the continuous magnetic field created by the sample sources. The spatial resolution that can be obtained from this 2D map of the field is then limited by the offset distance between the measurement plane and the sources, and by noise in the data \cite{Roth2D, MEG}. In general, the offset distance should be as small, or smaller than the characteristic length scale of the magnetic field sources, as shown in Fig.~\ref{BFields}, for reliable inversion of the magnetic image to the source distribution. In an analog to the Nyquist sampling theorem, the pixel size sets the maximum spatial (k-space) frequency. The field of view size sets the spatial frequency resolution (again by a Fourier transform argument). Both of these effects impact the ability to perform magnetic field inversions and map the underlying sources \cite{Roth2D}. 

\begin{figure*}[ht]
\begin{center}
\begin{overpic}[width=0.45\textwidth]{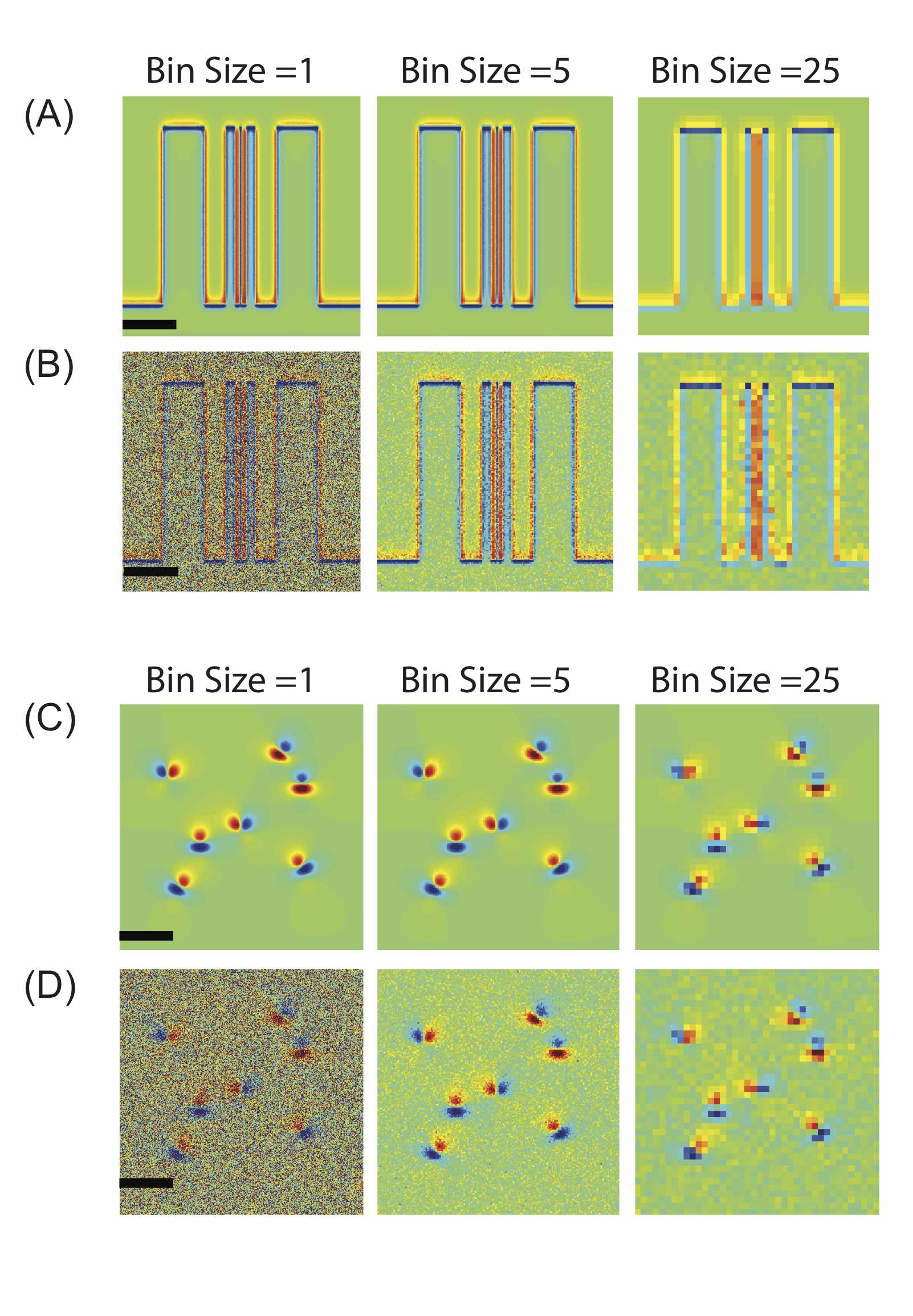}
\end{overpic}
\end{center}
\caption{\label{WireRes}
QDM Spatial Resolution and SNR Tradeoff. (A) Magnetic field from the current distribution in Fig.~\ref{BFields}(A) for different planar binning sizes. No additional noise is applied. Scale bar is 50 $\upmu$m (B) Binning with fractional noise leading to SNR of 1 for a bin size of 1. C)  Magnetic field from magnetic dipole sources in Fig.~\ref{BFields}(B). No additional noise is applied. Scale bar is 10 $\upmu$m (B) for different planar binning sizes (D) Binning with fractional noise leading to SNR of 1 for a bin size of 1. }
\end{figure*}

The in plane pixel size is made too small, then the noise level could preclude detection of magnetic fields of interest. This is similar to the negative impact to $\delta B$ that can result from pushing the temporal resolution, discussed in the previous section. On the other hand, if the pixel size is too large, then small length scale signals of interest will be blurred out and the fidelity of the magnetic field amplitude will be degraded. Fig.~\ref{WireRes} illustrates an example of this trade-off for magnetic fields simulated in Fig.~\ref{BFields}.

The QDM spatial resolution is set by the following:
\begin{enumerate}
    \item NV-sample standoff distance. As the standoff distance $\Delta z$ increases, the 2D magnetic map is convolved with a Lorentzian of width $\Delta z$, reducing the ability to resolve closely-separated magnetic sources \cite{eduardoUpcont}. Reducing the standoff distance improves the field strength and sometimes the spatial resolution.
    \item NV layer thickness. A thick NV layer has a layer-sample separation $\Delta z$ corresponding to somewhere between the NVs in the layer nearest to and farthest from the sample. An NV layer that is thin compared to the sample could have better sensitivity than with a thicker NV layer; the far-standoff NVs will measure a $\mathbf{B}_s$ comparable to that of the near-standoff NVs, and the photon shot noise improves with a thicker layer. Conversely, an NV layer that is thick compared to the sample will have far-standoff NVs that measure almost no field but add background fluorescence and can cause deleterious artifacts \cite{tetienneArtefacts}.
    \item Optical diffraction limit. set by the numerical aperture ($NA$) of the microscope objective  ($\lambda/(2 NA)$ for a typical fluorescence wavelength of $\lambda \approx 700$ nm. This assumes that the camera pixel size is small compared to the diffraction-limited spot size in the image plane. The spherical aberration from the diamond chip or other optics can also degrade the resolution. 
\end{enumerate}
QDM magnetic field imaging is best used for applications that need high spatial resolution over a wide-field of view, and can afford small NV-sample separation. The intuitive rule-of-thumb is to have NV layer thickness, standoff distance, and sample thickness of comparable sizes.

\section{QDM Components and Design Choices}\label{components}
The goal of an experimental designer is making sure that the application space of an experiment is limited by the fundamental physics of the system and not the equipment used. However, this is not always possible due to the availability and current state of technology. This section considers equipment choice and its impact on reaching the field sensitivity, temporal resolution, and spatial resolution prevented in Section 4 for different protocols.

Optimal performance for QDM target application can only be achieved with informed equipment choices.  These choices include proper selection of the diamond, bias magnetic field, MW field, optical illumination, optics, camera, and sample mounting. The QDM components and their impact on QDM performance is presented here with focus on informed hardware choices that enable operating the QDM at the physics-limited sensitivity and performance.

\subsection{Diamond}
Properties of the diamond chip that impact QDM performance include NV layer thickness, NV concentration, isotope and impurity concentration, and diamond cut. These properties are controlled for in the diamond fabrication process. Single-crystal diamond substrates used as the platform for QDM imaging are grown in one of two ways. One technique, high-pressure high-temperature (HPHT) growth, resembles natural diamond formation, requires an anvil press at $\sim$1700 K and 5 GPa, and produces diamond samples with $\sim$100 ppm nitrogen density. The second technique, chemical vapor deposition (CVD), grows diamond substrates layer-by-layer from a plasma, and yields diamond samples with low ppb nitrogen concentration.

Imaging a thin two-dimensional magnetic sample is optimal when the NV layer thickness is comparable to the magnetic sample thickness as discussed in Section 4.3. 

The typical NV layer thickness for QDMs ranges from $\sim$10 nm to $\sim$100 $\upmu$m. There are several methods available to make NV layers of varying thickness.
\begin{enumerate}
\item N$^+$ or N$_2^+$ is implanted in a type IIa diamond with ppb impurity density to create a $\sim$10-100 nm shallow layer. Annealing the diamond improves the NV yield and NV density \cite{HighTemp}.
\item A ppm-density nitrogen-rich layer is grown on top of a type IIa diamond substrate using CVD. After growth, electron-irradiation of the diamonds introduces vacancies, and annealing improves the NV yield by converting substitutional nitrogen atoms (P1 centers) into NVs with a $\sim$10$\%$ conversion rate \cite{Pezzagna_2010}. The nitrogen rich layer can be several microns down to several nanometers in thickness \cite{doi:10.1021/nl402286v}.
\item Instead of irradiating in $\#2$, the naturally-formed NVs are used and can be preferentially oriented along one of the crystallographic directions (instead of equal NV fractions along all four orientations). Removing three of the NV orientations can improve the magnetic field sensitivity by $\sim2\times$, but can come at the expense of reduced NV density and fluorescence \cite{1882-0786-10-4-045501}.
\item Similar to $\#2$, nitrogen is temporarily introduced during CVD diamond growth to create a few-nm nitrogen-rich layer, followed by a nitrogen-free diamond cap layer. NV centers are then created by electron-irradiating and annealing. This technique is called delta doping \cite{deltaDoping}. The cap layer adds to the standoff distance, so the surface-layer version in $\#2$ is often preferred, or the cap layer is etched away \cite{loretzAPL}.
\item An HPHT diamond with uniform NV volume density can be cut into $\sim$35 $\upmu$m thin slice. Alternatively, an HPHT diamond can be implanted with helium ions to form a shallow NV layer \cite{victor_mag, prepolNMR, victorHeImplant}.
\end{enumerate}

The NV density in the NV layer is optimized to achieve a desired magentic field sensitivity. High NV density yields more fluorescence intensity and good photon shot noise. However, the greater density of P1 paramagnetic impurities -- required for high NV yield -- contributes to magnetic inhomogeneity, thereby broadening ODMR resonances and spoiling magnetic field sensitivity. Optimal sensitivity therefore requires balancing the ODMR linewidth and contrast with the NV density in Eq.~\ref{SensivityODMR}. Conditions for a favorable ratio of the two NV charge states, NV$^-$ / NV$^0$, are also needed to ameliorate the NV$^0$ contribution to background fluorescence which spoils the NV$^-$ contrast used for imaging \cite{dianaChargeStates}.

The performance of diamonds with different C and N isotopes is an important consideration. The $^{15}$NV (spin-1/2 nucleus) is more optimal for QDM imaging because it gives greater ODMR contrast and requires a narrower range of MW probe frequencies than the more common $^{14}$NV (spin-1 nucleus). 
However, because $^{15}$N is the less abundant isotope, diamonds fabricated without special procedures for isotopic control will typically be dominated by $^{14}$N. 

Magnetic inhomogeneity from $^{13}$C (spin-1/2) and paramagentic P1 defect centers limits the NV $T_2^*$; thus, isotopically-purified $^{12}$C ($I=0$) diamonds are ideal \cite{P1DQ, longT2}. For diamonds with a 1.1$\%$ natural abundance of $^{13}$C present, it is advantageous to increase the P1 density resulting in larger NV density without contributing too much to the P1-limited $T_2^*$  \cite{victorHeImplant}. An NV layer fabricated in an isotopically-enriched $^{12}$C layer can reduce the ODMR linewidth. However, this may be irrelevant for NVs shallower than $\sim10$ nm, due to magnetic inhomogeneity introduced by electrons on the diamond surface \cite{marcAFMdeer}. 

Synthetic diamond chips used in QDMs are available in several cuts. The most common are diamonds with the top face along the [\textit{100}] plane and the sides along the [\textit{100}] or [\textit{110}] planes (Fig.~\ref{odmrExample}A). The NVs in these diamonds point roughly $35^{\circ}$ out of the plane. Less common diamond cuts include [\textit{110}] and [\textit{111}] top faces. The former has two NV orientations in-plane, while the latter has one NV orientation pointing normal to the face. Other more exotic diamond cuts exist, for instance Ref.~\cite{maletinskyMWimger} used a diamond with a [\textit{113}] NV layer. The choice of diamond cut does not usually impact the QDM performance. However, different cuts of diamond have different availability and pricing due to the challenge of producing crystals that are not grown along diamond's preferential growth axis. Surface termination effects can be of impact \cite{PhysRevLett.112.147602}.

The impact of diamond characteristics on specific QDM techniques is summarized as follows:
\begin{enumerate}
    \item For CW ODMR imaging, the laser and microwave linewidth broadening should match the diamond $T_2^*$ (\ref{ODMRSens}).
    \item For Ramsey imaging, the diamond $T_2^*$ limits the phase accumulation time.
    \item For dynamical decoupling imaging, the diamond $T_2$ limits the phase accumulation time (depending on the magnetic noise spectrum and pulse sequence).
    \item For Rabi and $T_1$ imaging, the diamond $T_2^*$ sets the spectral filter function. The intrinsic NV $T_1$ depends on the NV density and depth.
\end{enumerate}

\subsection{Laser}
A QDM typically uses a 532 nm solid-state laser for optical pumping due to availability and performance. The green pump laser intensity is weak, typically $\sim$10 - 1,000 W/cm$^2$, when illuminating a few-mm field of view, which can be a limitation for pulsed-readout techniques. The NV $^3A_2 \rightarrow {^3}E$ optical transition spans hundreds of nanometers wavelength due to the phonon sideband as discussed in section 2.2, which allows for laser excitation wavelengths ranging from 637 nm to $\sim$ 470 nm \cite{Kilin2000201}. Past experiments have pumped the NVs with 532 nm frequency-doubled Nd:YAG and Nd:YVO$_4$ lasers, 637 nm and 520 nm diode lasers and LEDs, 594 nm HeNe lasers, argon-ion laser lines (457, 476, 488, 496, and 514 nm), and  supercontinuum lasers with an acousto-optic tunable filter \cite{wrachtrup_darkstates, erikThesis, LEDmagnetometer}. There have been attempts to find the illumination wavelength with the most favorable cross-section and NV$^-$/NV$^0$ charge-state ratio \cite{optimum_phot}. Since the NV readout measures a fluorescence intensity, fluorescence intensity instability from the laser or the optics must be minimized for the QDM magnetic sensitivity to reach the photon shot-noise limit.

Increasing the illumination intensity improves the NV fluorescence intensity, the photon shot noise, and sometimes the ODMR lineshape. The $^3A_2 \rightarrow {^3}E$ optical transition is dipole-allowed when illuminating with light polarized in the ${x,y}$ plane of the NV coordinate system defined in section 2.1 \cite{victor_singlets}. Thus, in a projection magnetic microscopy experiment (Fig.~\ref{odmrExample}C) a laser polarization is chosen that favors the optical absorption selection rules for the selected NV orientation. If all NV orientations are interrogated, a laser polarization is selected that addresses all NV orientations with comparable strength. Increasing the laser illumination power increases the diamond and sample heating on approximately linear scaling, while the photon-shot-noise limit only increases as the square root of the laser power. Furthermore, as the fractional photon shot noise improves, the analog-to-digital conversion bit depth must also improve to avoid being quantization-noise-limited.

The available laser intensity affects the various QDM techniques in the following ways:
\begin{enumerate}
    \item For CW ODMR imaging, varying the laser intensity affects the ODMR linewidth (Fig.~\ref{ODMRSens}). 
    \item For pulsed imaging experiments, ideally the laser intensity should be close to optical saturation. Weaker laser intensity, longer $t_I$, and longer $t_R$ will worsen the experiment time resolution.
\end{enumerate}{}

\subsection{Microwave Source}
The simplest way to apply a MW field to the NVs is with a piece of wire connected to a coaxial cable. The QDM MW field is ideally uniform across the NV layer field of view, and there are a variety of alternative engineered MW antennas that aim to optimize the MW field homogeneity, efficiency, or bandwidth \cite{loncarSplitRing, Zhang2016, UCBferromagMW, ultrabroadbandCPW, itohMWresonator, llBroadbandLoopGap, polandCircPol}.  By the transition selection rules, the transitions between $^3A_2$ sublevels require left-circularly or right-circularly polarized MW  \cite{HPpolDependence}. One QDM MW antenna option is a pair MW loop as shown in Figure \ref{Fig1}; another option is a of crossed  MW stripline resonators \cite{QDM1ggg}. The striplines are excited in-phase (or 90 degrees out-of-phase) to produce a linearly (or circularly) polarized MW field as needed for a given sensing protocol.

Increasing the MW power improves the contrast in a CW ODMR measurement, but also broadens the linewidth between $^3A_2$ resonances. Fig.~\ref{ODMRSens} demonstrates for simulated CW ODMR measurements that optimizing QDM magnetic sensitivity implies tradeoffs of laser and MW power \cite{dreau_powBroad}. Choosing an optical pumping rate much greater than the MW transition rate results in weak contrast, since the laser quickly repumps any NV population fraction removed by resonant MWs. Increasing the MW field amplitude  improves the fluorescence contrast but also broadens the ODMR linewidth. The MW intensity noise also affects the QDM sensitivity by influencing the ODMR contrast and linewidth in a manner similar to fluorescence intensity noise.

Selecting an appropriate MW frequency sweep rate is critical in CW ODMR measurements where we sweep the probe MW frequency across the NV resonance. The NV reaction time depends on the NV optical pumping rate and MW transition rate, and a sufficient response time is needed for the NVs to re-equilibrate to the updated conditions after updating in the MW probe frequency. This also applies to experiments using lock-in detection to combat fluorescence intensity noise: the MW modulation rate must be slower than the NV reaction time, typically set by the optically pumping rate \cite{shin:124519}. 

When deciding how to apply the microwave field, some of the options affect the various QDM modalities differently:
\begin{enumerate}
    \item For CW ODMR imaging, increasing the MW power broadens the ODMR linewidth but also improves the contrast.
    \item For Ramsey, Hahn Echo, and dynamical decoupling imaging, spatial MW inhomogeneity and pulse errors can reduce the NV contrast and worsen the sensitivity.
\end{enumerate}

\subsection{Static Magnetic Field}
The QDM bias magnetic field $\mathbf{B}_0$ can be provided by electromagnets (Helmholtz coil sets, solenoids, and C-frame/H-frame electromagnets) or permanent magnets \cite{QDM1ggg, liamT1, Glenn2018} as shown in Figure \ref{Fig1}. Electromagnets allow us to select arbitrary $|\mathbf{B}_0|$ up to a few tesla. However, they require a stable current supply, may need water cooling for the magnet, and can add to sample and system heating. Permanent magnets allow higher $\mathbf{B}_0$ in a more compact instrument, though the applied $\mathbf{B}_0$ can drift with temperature. 

\begin{figure*}[ht]
\begin{center}
\begin{overpic}[width=0.55\textwidth]{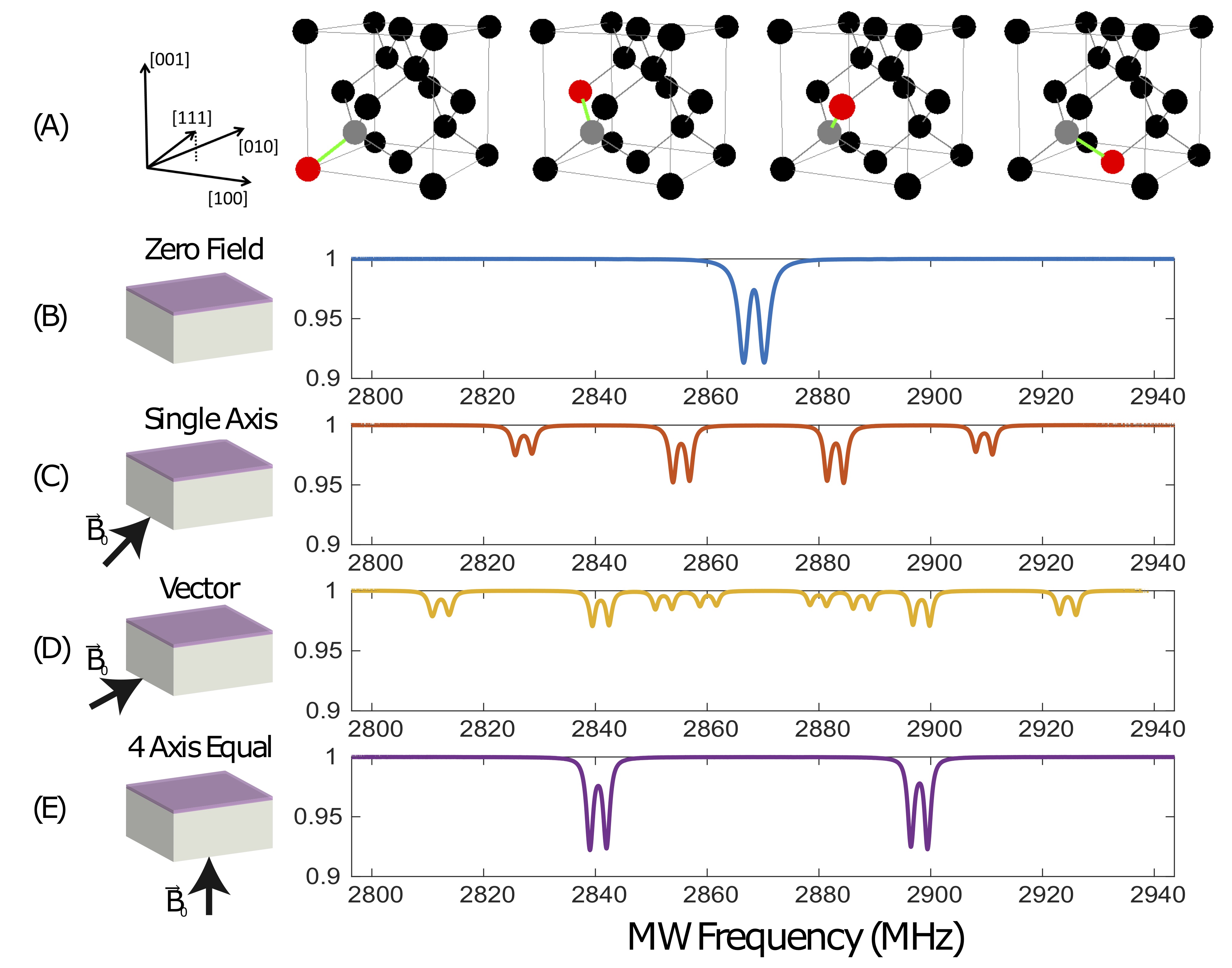}
\end{overpic}
\end{center}
\caption{\label{odmrExample} Experimental ODMR Spectra for Different Bias Magnetic Field Magnitudes and Orientations.
(A) Example of four possible NV orientations in the diamond lattice, and the crystallographic directions.
(B) Example ensemble NV ODMR spectrum with $|\mathbf{B}_0|=0$. The resonance is centered at $\sim$2.87 GHz, but splits into two peaks around this resonance frequency due to the $^{15}$N hyperfine coupling. Strain and electric field also contribute to the ODMR lineshape and broadening, and can cause a variety of lineshapes at $|\mathbf{B}_0|=0$ for different samples.
(C) Ensemble NV ODMR spectrum with $|\mathbf{B}_0|$ pointing along one axis. The frequency separation between the outer resonance peaks is proportional to the applied field. The inner peaks are from the three other NV orientations overlapping with each other due to equal Zeeman interactions for each. The $^{15}$N hyperfine interaction again splits each resonance into a doublet.
(D) Ensemble NV ODMR spectrum with $|\mathbf{B}_0|$ orientation such that each axis has different projection of bias field. 
(E) Ensemble NV ODMR spectrum with $|\mathbf{B}_0|$ along the [001] direction, such that each NV orientation has the same Zeeman interaction. The peak separation is proportional to the $|\mathbf{B}_0|$ field projection along the NV axes.}
\end{figure*}

The choice of bias field amplitude $|\mathbf{B}_0|$ depends on the samples being measured. Soft magnetic samples that might have their magnetization changed by an applied magnetic field require $|\mathbf{B}_0|$ to be minimized. This has the added benefit that small $|\mathbf{B}_0|$ typically implies a small $|\mathbf{B}_0|$ gradient  across the field of view. A large $|\mathbf{B}_0|$ can be beneficial when imaging paramagnetic minerals, since the magnetization from paramagnetic particles scales with $|\mathbf{B}_0|$ until saturation \cite{hemozoin}. For Rabi imaging or $T_1$ magnetometry, $|\mathbf{B}_0|$ is chosen such that the NV spin transition frequency matches the AC sample frequency being interrogated \cite{maletinskyMWimger, liamT1}. Due to nitrogen nuclear polarization, $|\mathbf{B}_0|$, $\sim$30-50 mT improves the NV fluorescence contrast \cite{ransPaper, steiner_hf}. Finally, $|\mathbf{B}_0|=0$ is an intuitive choice for NV thermometry or electrometry experiments (Fig.~\ref{odmrExample}B)

The direction of $\mathbf{B}_0$ also factors into the specific QDM application \cite{QDM1ggg}. Alignment of $\mathbf{B}_0$ along the N-V axis ([\textit{111}] crystallographic direction) allow for interrogating the NVs along this direction, Fig.~\ref{odmrExample}C. This approach allows optimization of the other measurement parameters, e.g., the optical polarization, to maximize the fluorescence and contrast from the selected NV orientation. Alternatively the $\mathbf{B}_0$ magnitude and direction are chosen such that each NV orientation has different resonance frequencies and non-overlapping spectra, Fig.~\ref{odmrExample}D. This approach allows reconstruction of vector magnetic field information from the eight NV resonance frequencies. $\mathbf{B}_0$ can also be aligned with the crystallographic [100] or [110] directions, such that the resonance frequencies for different NV orientations are degenerate, leading to improved contrast. If $\mathbf{B}_0$ is aligned along the [100] direction with a diamond cut along [100], the magnetic field projection direction is normal to the chip, though the Zeeman shift is $\sqrt{3}$ times weaker than for $\mathbf{B}_0$ along the [111] direction (Fig.~\ref{odmrExample}E). Finally, there may be some experiments where the choice of $\mathbf{B}_0$ is forced by the sample being tested. This could cause the NV ODMR lines to overlap, making it difficult to resolve the resonance frequencies and the extract vector magnetic field information. This difficulty can be ameliorated using the $^3A_2 \leftrightarrow ^3E$ optical polarization selection rules to distinguish the light contributions from each NV orientation \cite{mikaelFourier}.

The $\mathbf{B}_0$ field is ideally as uniform as possible. $\mathbf{B}_0$ inhomogeneity can cause the following problems:
\begin{enumerate}
    \item In pulsed NV experiments, $\mathbf{B}_0$ inhomogeneity will cause spatially-dependent pulse errors, which limit the NV contrast and sensitivity.
    \item For all experiments, a $\mathbf{B}_0$ gradient on top of a the desired $\mathbf{B}_s$ is something that should be subtracted out. A uniform $\mathbf{B}_0$ allows for subtraction of a constant offset. 
    \item In an extreme case, $\mathbf{B}_0$ inhomogeneity can contribute to NV linewidth broadening within each pixel.
\end{enumerate}

\subsection{Optics}

QDMs employ various ways to illuminate the NV layer with pump-laser light, depending on other experimental constraints. Side illumination of the diamond chip \cite{QDM1ggg} is a good method for QDMs using a low-magnification (long working distance) objective with a large field of view since the beam will have enough space to avoid clipping the objective and also illuminate a large area. Another approach is to illuminate  through the objective by focusing the pump laser at the back aperture to get parallel rays out of the objective \cite{nirPaleomag}. This method works better for QDMs operating with high-magnification microscopes. The laser polarization is easier to control, but focusing the laser at the objective back aperture can lead to burns. Techniques to avoid illuminating the sample as well as the NVs include illumination via total internal reflection in the diamond, shaping the pump laser beam into a light sheet using cylindrical lenses, or coating the NV surface with a reflective layer to reduce the optical intensity through the diamond chip \cite{nv_bacteria, maletinskyMWimger, barryNeurons}. 

Optimal photon collection efficiency requires the largest achievable numerical aperture (NA) for the microscope objective. In practice the NA for a given magnification is limited, and high-NA objectives are often also high-magnification objectives with a short working distance (sometimes shorter than the diamond thickness). Imaging NV fluorescence through the diamond chip may cause optical aberrations that can spoil the image quality, though we are unaware of any QDM experiment that corrects for this. As with other optical microscopes, a QDM images a broadband NV fluorescence ($\sim$637-800 nm), so chromatic aberration in the microscope optics is also important to mitigate. Pulsed NV experiments commonly use an acousto-optic modulator (AOM) as an optical switch. For the AOM, the rise-time, extinction ratio, and efficiency are the parameters to consider for a given application.

\subsection{Camera}

QDM camera selection for a targeted application requires consideration of the expected photon collection rate from the NV layer, camera read noise and dark-current noise, well depth, global/rolling shutter capability, software and external triggering, frame rate, data transfer rate, pixel size, and quantum efficiency \cite{danishCameraComp}. For experiments with a high photon count rate, the camera must handle enough photoelectrons per second without saturating. Here, the pixel well depth, number of pixels, quantum efficiency, and frame rate are the important quantities to consider, because they determine the maximum photon count rate for fluorescence detection. 

The camera frame rate can limit the experimentally-realizable temporal resolution. Increasing the camera frame rate is possible by using only a fraction of the sensor. However the resulting product between the frame rate and the number of pixels usually decreases, indicating that use the full camera sensor is better for maximizing the number of photoelectrons per second. Alternatively, if the photon count rate is low, parameters like the read noise and dark-current noise should be minimized while the quantum efficiency is maximized. For pulsed experiments, a slow camera frame rate can throttle the experiment repetition rate and sensitivity.

The camera sensor size determines the microscope magnification for a desired field of view. The microscope spatial resolution can be set by the camera pixel size (rather than the optical diffraction limit) if the camera pixels are too widely spaced for the microscope magnification. The diffraction-limited spatial resolution should be oversampled by at least $2\times$ to avoid having the pixel size spoil the diffraction-limited spatial resolution. A given choice of microscope optics has a finite effective image area, and the camera image can have darkened corners (vignetting) if the camera sensor area is too large.

As previously mentioned, the optical readout time, $t_R$, must be balanced with the minimum camera exposure time, and the maximum camera frame rate for pulsed QDM experiments. Specifically:
\begin{enumerate}
    \item Sensitivity is lost for experiments with a measurement time, $t_{meas}$, faster than the camera frame rate, because the camera is too slow to acquire a new frame at the rate it takes to do each experiment.
    \item Experiments for which the minimum camera exposure time is longer than $t_R$ require the readout laser to be off for the duration of the time difference.
\end{enumerate}

\subsection{Diamond Mounting and Configuration}
There are two primary ways to prepare the diamond sensor chip and the sample in the QDM. The first method is to fix the diamond chip in the optical microscope setup and move the sample independently with kinematic stages. This way the diamond chip position (and all other optics) are permanent, keeping the relative positions of the optics, diamond location and orientation, MW field, and magnetic field constant for all measurements to improve reproducibility and enable faster setup time for new samples. The second method is to mount the diamond chip directly on the sample, then move the diamond and the sample together within the microscope field of view. This  integrated diamond/sample approach offers more certainty that the NV-sample separation is minimized. Generally, sample mounting and manipulation in a QDM is easier with an upright microscope setup rather than an inverted microscope. 

\begin{table}[]
\resizebox{\textwidth}{!}{%
\begin{tabular}{|c|l|l|l|l|}
\hline
\textbf{Design choice}  & \multicolumn{1}{c|}{\textbf{Diamond}} & \multicolumn{1}{c|}{\textbf{Laser}} & \multicolumn{1}{c|}{\textbf{Microwaves}} & \multicolumn{1}{c|}{\textbf{$\mathbf{B}_0$ field}} \\ \hline
\textbf{Considerations} & \begin{tabular}[c]{@{}l@{}}- NV density affects the \\ sensitivity\\ - Inhomogeneity in strain,\\ density, and magnetic \\ environment spoils the \\ sensitivity\\ - Match the NV layer \\ thickness and sample \\ thickness\end{tabular} & \begin{tabular}[c]{@{}l@{}}- field of view sets laser\\ intensity\\ - Laser intensity noise can \\ limit sensitivity\\ - Laser polarization addresses \\ different NV orientations\\ - Homogeneous illumination\\ is desirable\end{tabular} & \begin{tabular}[c]{@{}l@{}}- Amplitude and phase \\ instability affect sensitivity. \\ - Amplitude homogeneity \\ is desirable\end{tabular} & \begin{tabular}[c]{@{}l@{}}- A $\mathbf{B}_0$  gradient can cause \\ varying sensitivity uniformity\end{tabular} \\ \hline
\textbf{Design choice}  & \multicolumn{1}{c|}{\textbf{Optics}} & \multicolumn{1}{c|}{\textbf{Camera}} & \multicolumn{1}{c|}{\textbf{Diamond mount}} & \multicolumn{1}{c|}{\textbf{Magnetic environment}} \\ \hline
\textbf{Considerations} & \begin{tabular}[c]{@{}l@{}}- Microscope objective\\ sets the collection \\ efficiency and optical \\ diffraction limit\\ - Microscope components \\ set the magnification and \\ field of view size\end{tabular} & \begin{tabular}[c]{@{}l@{}}- Pixel size should oversample \\ other resolution limitations \\ (e.g. optical diffraction)\\ - Frame rate x well depth x \\ number of pixels set the \\ best-possible sensitivity\\ - Transfer rate and buffer size \\ can throttle the maximum \\ experiment rate\\ - Camera efficiency is worse \\ than photodiode efficiency\end{tabular} & \begin{tabular}[c]{@{}l@{}}- Aim for high thermal and \\ mechanical stability during \\ an experiment\end{tabular} & \begin{tabular}[c]{@{}l@{}}- Mitigate background field\\ (e.g. Earth's field, electronics, \\ ...)\end{tabular} \\ \hline
\end{tabular} }
\label{tblSec5gen}
\caption{General QDM hardware considerations that apply to all measurement techniques}
\end{table}

\subsection{General Design Considerations}
Table \ref{tblSec5gen} summarizes equipment parameters that optimize QDM operation. While some of the above specifications are technique- or application-specific, this table describes general design choices that affect all QDM instruments.

\section{QDM Applications}
QDM magnetic field imaging has been applied to a diverse  range  of applications  across  numerous  fields  of  research. For every given application, the appropriate experimental protocol must be chosen for optimal performance, including the desired temporal resolution and magnetic frequency range. This subsequently dictates the QDM component implementation. Table \ref{QDMApplications} lists the application target areas for the various QDM techniques and respective frequencies. To more easily motivate future unrealized QDM applications, the following sections includes examples of successful QDM applications for each frequency range and application area. 

\begin{table}[]
\resizebox{\textwidth}{!}{%
\begin{tabular}{|c|c|c|c|}
\hline
             & Broadband 0-1MHz                                                                                                        & Narrowband $\sim$1kHz-20MHz                                                                            & Narrowband 10MHz-100GHz                                                                                               \\ \hline
Techniques   & \begin{tabular}[c]{@{}c@{}}CW ODMR\\ Pulsed ODMR\\ Ramsey\end{tabular}                                                    & \begin{tabular}[c]{@{}c@{}}Hahn Echo\\ Dynamical Decoupling\end{tabular}                          & \begin{tabular}[c]{@{}c@{}}CW ODMR\\ Rabi\\ $T_1$ Relaxation\end{tabular}                                             \\ \hline
Applications & \begin{tabular}[c]{@{}c@{}}Paleomagnetism and Rock magnetism\\ Biomagnetism\\ Solid State Magnetism\\ Low Frequency Electronics\end{tabular} & \begin{tabular}[c]{@{}c@{}}RF Electronics\\ Solid State Magnetism\\ NMR Spectroscopy\end{tabular} & \begin{tabular}[c]{@{}c@{}}MW Electronics\\ Solid State Magnetism\\ EPR Spectroscopy\\ NMR Spectroscopy\end{tabular} \\ \hline
\end{tabular}%
}
\caption{\label{QDMApplications} QDM Techniques and Applications. Overview of techniques and potential applications for widefield magnetic imaging.}
\end{table}

\subsection{Broadband Imaging of 0-1 MHz Magnetic Fields}

\begin{figure*}[ht]
\begin{center}
\begin{overpic}[width=0.75\textwidth]{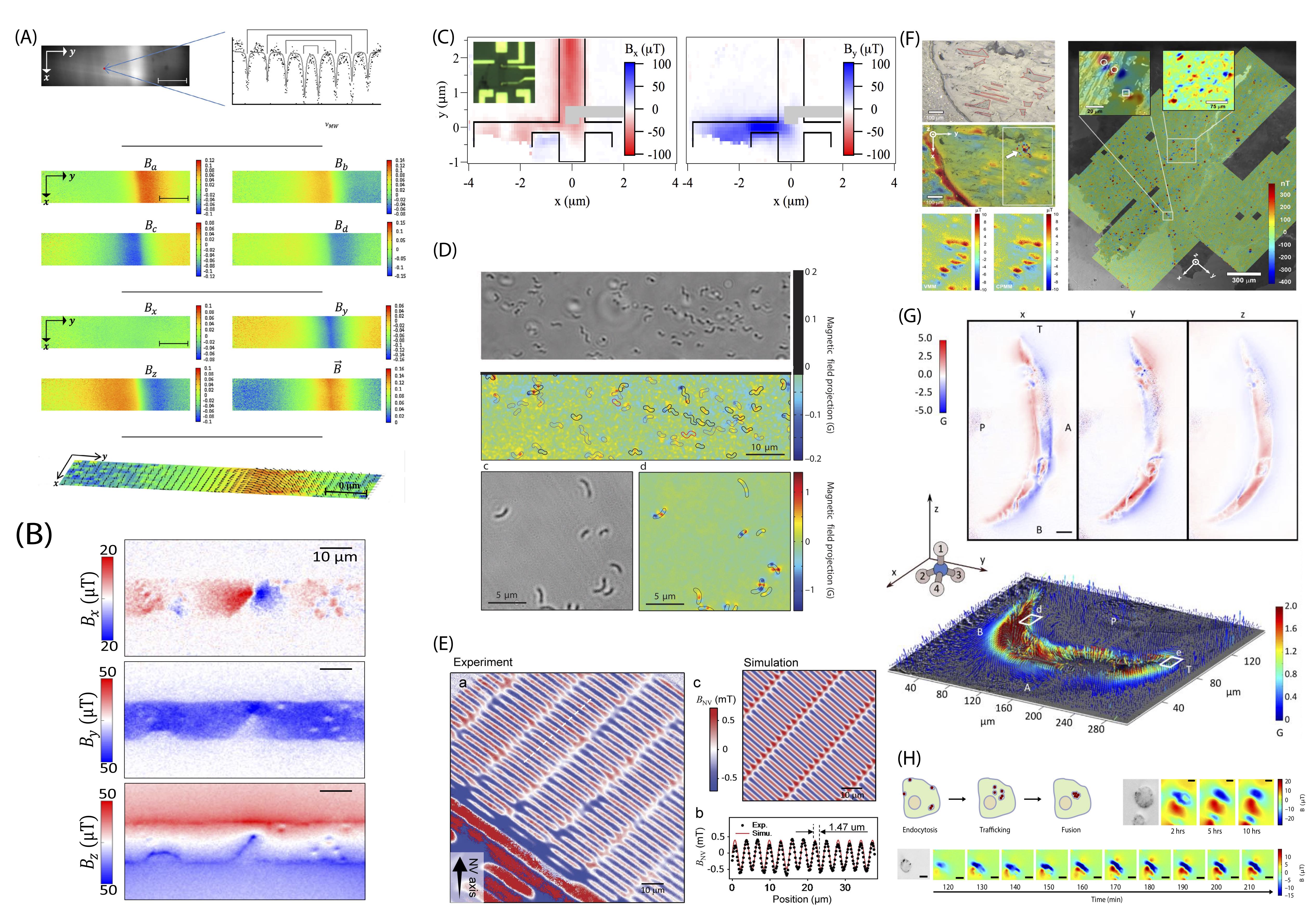}
\end{overpic}
\end{center}
\caption{\label{DCExamples}
Examples of QDM DC Magnetic Imaging. (A) Imaging the vector magnetic field from a wire on the diamond. Panel A reprinted with permission from Ref \cite{roch_magImg}. (B) Example of magnetic field and reconstructed current from current flow in graphene \cite{Tetiennee1602429} (C) Image showing parabolic profile of hydrodynamic flow in graphene at the Dirac point \cite{DiracGraphene} (D) Magnetic field image of magnetite intrinsic to MTB bacteria. Panel D reprinted with permission from Ref \cite{nv_bacteria}. (E) Imaging static magnetic field profile associated with magnetic memory \cite{hollenbergFilms} (F) Measurement of remnant magnetization from geological sample. Panel F reprinted with permission from Ref \cite{QDM1ggg}.  (G) Imaging magnetic field from iron mineralization in chiton teeth \cite{ChitonTeeth}  (H) Visualization of trafficking of magnetite particles in biological tissue \cite{shapiroMRI}}
\end{figure*}

CW ODMR imaging experiments of static magnetic fields is among the most successful QDM imaging to date. The relatively low MW and laser power requirements and simplicity of the experimental control allows for imaging of static magnetic fields up to a 4$\times$4 mm region (limited by the size of a diamond substrate.) Most experiments up to this point have chosen to focus on large quasi-static magnetic fields due to the relatively loose requirements for performance of the QDM.  Figure \ref{DCExamples} shows several examples.

\subsubsection{Current Distributions}

Imaging magnetic fields from 2D current distributions was among the first demonstrations of a QDM system due to the flexibility in choice of magnetic field amplitude, temporal profile, and spatial structure, making it well suited for verifying the fidelity of magnetic field imaging experiments \cite{imaging_mag, roch_magImg}. For sufficiently simple wire patterns, one can simulate the expected current distribution and field map to compare with and validate the QDM measurement. In these early experiments the large sample current amplitudes were needed because of limited diamond  sensitivity values at the time. 

Magnetic field imaging for determining current flow along circuit traces was one of the first demonstrations \cite{Circuit} of imaging the vector component from a nontrivial current distribution. However, the sensitivity of the system was not optimized and nontrivial temporal dynamics of current flow were not investigated. 

Magnetic field imaging can allow for the interrogation of nontrivial current flow in 2D materials. Magnetic imaging has been utilized in probing the spatial dynamics of current flow in graphene. In an initial demonstration, current was passed through graphene and defects in the 2D graphene were apparent due to the current flowing around them \cite{Tetiennee1602429}. In a separate experiment, magnetic imaging experiments were performed to probe the viscous Dirac fluid nature of current near the Dirac point. High resolution magnetic field imaging allowed for a direct measurement of the parabolic current profile associated with the hydrodynamic behavior of this Dirac fluid \cite{DiracGraphene}.

Looking to the future, the application space can dramatically improve if Ramsey Imaging is implemented and optimized. With an optimized version of a Ramsey Imaging system, there is projected to be sufficient volume normalized sensitivity to image the propagation of activity associated current in a mammalian neuron in real time \cite{barryNeurons}.

\subsubsection{Magnetic Particles and Domains}

Measuring the DC component of magnetic particles and domains has yielded some of the most transformative applications of widefield magnetic imaging to date. Examples in the literature span from magnetotactic bacteria \cite{nv_bacteria} and magnetically-labeled cells \cite{glennCancer, GouldLabel} to remanent magnetization in geological samples \cite{QDM1ggg} and thin magnetic films \cite{hollenbergFilms}. Success in these applications has been due in part to the generally static (enabling signal averaging) and large magnetic fields produced by these sources which together reduce the need to push the state of the art on sensitivity.

In the earliest biological QDM experiment, the intrinsic magnetite inside magnetotactic bacteria was measured \cite{nv_bacteria}, as shown in Fig.~\ref{DCExamples}D. Other work has been performed to look at the intrinsic magnetite in chiton teeth to study iron mineralization \cite{ChitonTeeth} (see Fig.~\ref{DCExamples}G) and malarial hemozoin nanocrystals \cite{hemozoin}.

Magnetically labelling cells is a promising technique for tracking and identifying rare cell types \cite{glennCancer, GouldLabel}. Other groups have followed up on this work with extrinsic magnetic particles in applications relating to probing the origin of contrast agents in MRI \cite{shapiroMRI}, as shown in Fig.~\ref{DCExamples}H, and furthering the imaging resolution and sensitivity on magnetic particle imaging \cite{Microbeads,doi:10.1021/acs.nanolett.8b03222}.

QDMs have proven to be a valuable component in the toolbox of remanent magnetization studies in geological samples (see Fig.~\ref{DCExamples}F) \cite{QDM1ggg}. Initial demonstrations \cite{Fu1089} were performed on the Semarkona meteorite to assist in determining the magnetic field present during planetary formation. Followup work with the QDMs have demonstrated their utility in imaging magnetization carriers at the grain scale. Recent example applications have included the imaging of large magnetite grains to visualize multi-domain structure \cite{nirPaleomag} and of zircons \cite{FU20171, Tang407, dynamo} to understand and constrain the history of Earth's dynamo. The full potential of the QDM as a rock magnetic instrument are only beginning to be explored, with ongoing experiments on terrestrial and extraterrestrial rock types being pursued. 

QDMs have extended their range to condensed matter to probe thin magnetic films such as magnetic memory (see Fig.~\ref{DCExamples}E) \cite{hollenbergFilms} and explore questions related to the origins and properties of vortices in superconductors \cite{heziVortices}.

\subsection{Narrow-band Imaging of $\sim$1 kHz - 20 MHz Magnetic Fields}

\begin{figure*}[ht]
\begin{center}
\begin{overpic}[width=0.75\textwidth]{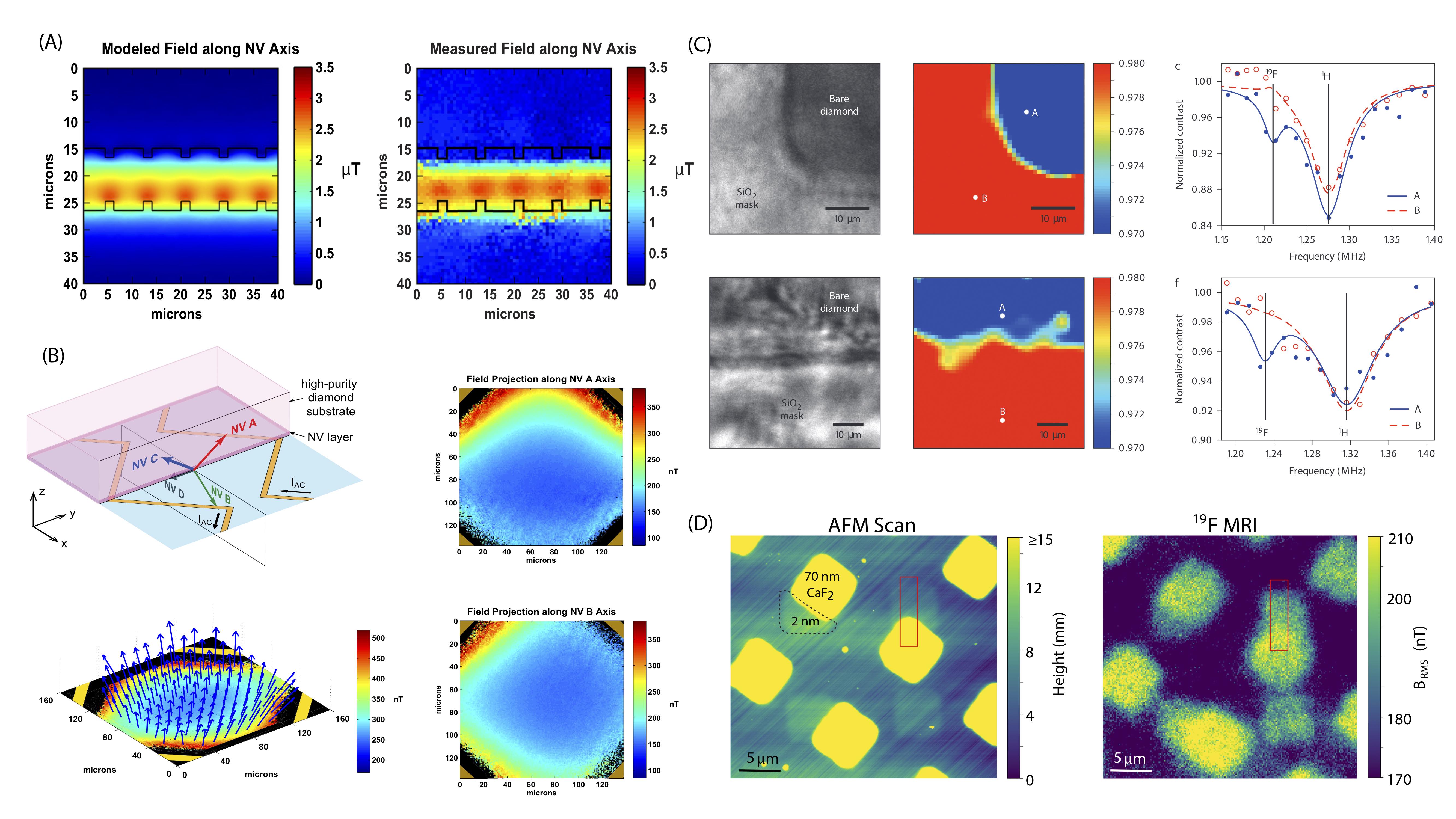}
\end{overpic}
\end{center}
\caption{\label{ACExamples}
Examples of Low-Frequency QDM AC Magnetic Imaging. (A) Imaging magnetic field from current oscillating  at 109.5 kHz \cite{linh_widefield}. (B) Imaging magnetic field from a wider field of view of current oscillating at 4.75 kHz \cite{linh_widefield}. (C) Imaging presence of $^{19}$F on diamond surface through NMR signal. Panel reprinted with permission from \cite{devience_nuclSens} (D) High spatial resolution imaging of patterned $^{19}$F on diamond surface \cite{1807.08343}}
\end{figure*}

Narrowband magnetic imaging in an intermediate frequency range is mostly applicable for imaging magnetic fields originating from current distributions and the magnetic field from precessing spins in nuclear magnetic resonance (NMR) applications, shown in Figure \ref{ACExamples}. Much of the early work in NV magnetic imaging pushed the state of the art in these regimes, but more development is needed to explore the full range of applications. 

\subsubsection{Current Distributions}

Similarly to the broadband case, current distributions were initially used to validate the fidelity and effectiveness of AC magnetometry pulses sequences in an imaging modality \cite{linh_widefield}. In this demonstration current with frequencies ranging from 4 kHz to 100 kHz were sent through wires fabricated on the diamond. 

One promising application of this technique is imaging magnetic fields which oscillate near the clock frequency of circuits for side channel attack analysis \cite{10.1007/3-540-36400-5_4}. NV Diamond can allow for the ability to include spatial information. Extending the sensing frequency beyond $\sim$20 MHz is challenging for several reasons. For sensing high frequencies, the MW $\pi$-pulse duration should be short compared to the period of the sensing signal. Short $\pi$ pulses require strong MW fields to achieve the high Rabi frequencies. Strong, uniform MW pulses over a large area are a difficult engineering challenge and requires more sophisticated MW antenna design. Furthermore, even if these requirements are fulfilled, the strong MW fields can interfere or damage the sample being sensed. 

\subsubsection{NMR Signals}

Nuclear magnetic resonance (NMR) spectroscopy allows label-free detection and quantification of molecules with excellent chemical specifity. The use of narrowband AC magnetic imaging techniques to record local NMR signals in individual QDM pixels opens the possibility of highly-multiplexed, two-dimensional spatial density mapping of arbitrary molecular species. Potential applications include imaging small-molecule concentrations in neuronal slice preparations or bacterial biofilms \cite{NMRMatrix}; spatially-resolved battery electrochemistry \cite{NMRBattery}; detection and determination of the chemical composition of proteins \cite{igorProteinNMR}; or possibly a readout for molecular data storage \cite{NMRInformation}. 

Even without the high spectral resolution required to distinguish molecular species (typically ~1 ppm of the nuclear Larmor frequency or better, which  places stringent technical constraints on the magnitude, stability and homogeneity of the bias magnetic field $\mathbf{B}_0$), the combination of QDM imaging with correlation spectroscopy techniques \cite{13CCorr} and/or strong pulsed magnetic gradients \cite{keigo_fourier} can provide spatial maps of sample physical properties such as fluid density, net flow velocity fields, and/or local diffusion rates \cite{ncomms9527}. This could have applications to the study of porous media in petrochemistry, filtration, or catalysis.

As with broadband Ramsey spectroscopy, the pulse sequences used for narrowband AC magnetic imaging necessitate efficient temporal segmentation of NV fluorescence data at timescales of $<$1 microsecond, which is challenging for standard scientific imaging cameras. For this reason, there have been few reported demonstrations of NMR signal imaging using QDMs reported in the literature to date \cite{devience_nuclSens, 1807.08343}, and none with the spectral resolution required to distinguish molecular species. Nevertheless, we anticipate that ongoing work to integrate broadband Ramsey spectroscopy into the QDM platform can be directly extended to narrowband AC signal detection, and ultimately to high spectral-resolution NMR readout techniques \cite{Glenn2018, Schmitt832,Boss837, 1807.04444}.

\subsection{Narrow-band Imaging of 10 MHz - 100 GHz Magnetic Fields}

\begin{figure*}[ht]
\begin{center}
\begin{overpic}[width=0.75\textwidth]{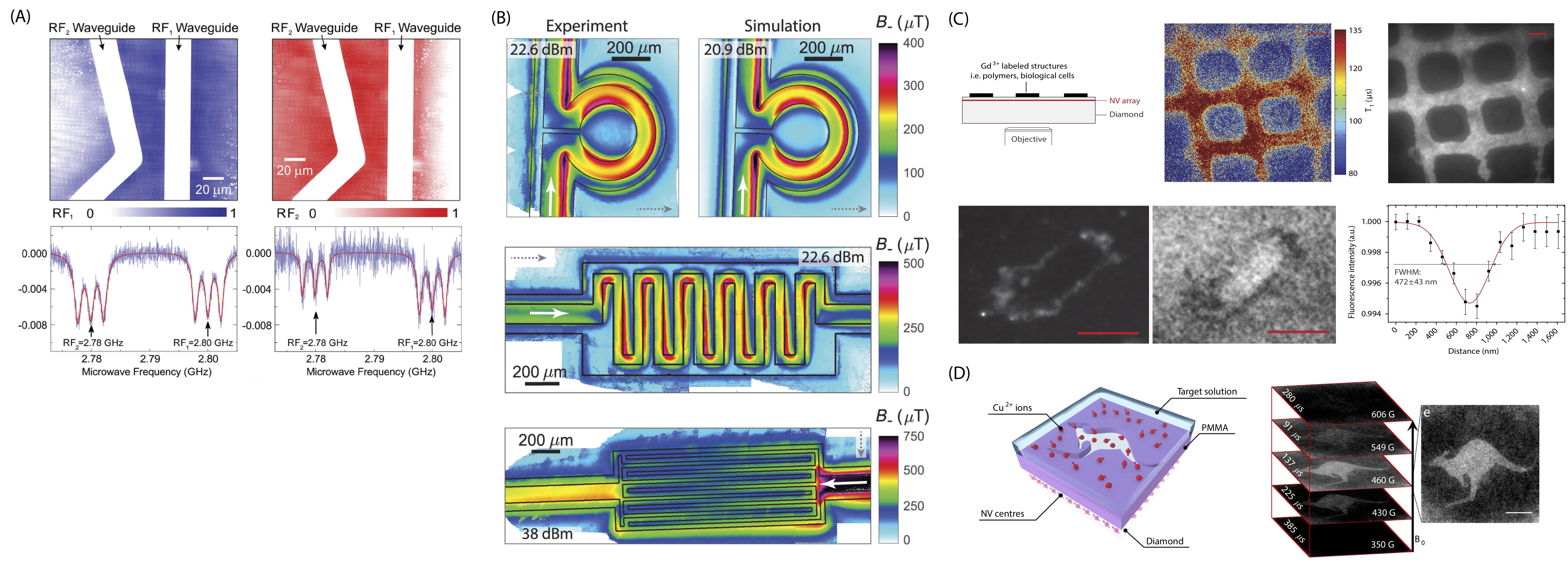}
\end{overpic}
\end{center}
\caption{\label{GHzExamples}
Examples of  GHz-Frequency QDM AC Magnetic Imaging. (A) Imaging presence of MW field through the contrast and linewidth of the ODMR. Panel A reprinted with permission from Ref \cite{linboMWimging}. (B) High spatial and temporal resolution Rabi Imaging. Panel B reprinted with permission from Ref \cite{maletinskyMWimger}. (C) $T_1$ weight imaging of patterned Gd$^{3+}$ and a Gd$^{3+}$ labeled cell membrane. Panel C reprinted with permission from Ref \cite{T1flowimaging}. (D) Demonstration of $T_1$ imaging of patterned Cu$^{2+}$ ions and sensitivity to the bias magnetic field \cite{EPRSimpson}}
\end{figure*}

\subsubsection{Microwave Imaging}
    QDM imaging of  the microwave field from wires, resonators, and structures is possible by measuring the Rabi frequency in a pulsed experiment \cite{maletinskyMWimger}, or by using the fluorescence contrast in a CW ODMR experiment  \cite{linboMWimging} (see Figure \ref{GHzExamples}A and \ref{GHzExamples}B). An initial step is to  compare the NV measurement to a predicted magnetic field map from a finite element method (FEM) calculation. One  goal is to use NV microwave imaging as a tool to validate that the FEM or the fabrication are what expected for more nontrivial devices like atom chips.

\subsubsection{$T_1$ Imaging for Paramagnetic Spins}
Just as coherent resonant microwaves drive transitions between the NV $^3A_2$ sublevels,  external paramagnetic spins can have the same effect, spoiling the NV $T_1$ (see Figure \ref{GHzExamples}C and \ref{GHzExamples}D). Paramagnetic spins with short $T_1$ can produce broadband magnetic noise that spoils the NV $T_1$, while long-lived paramagnetic spins can spoil the NV $T_1$ for particular $|\mathbf{B}_0|$ where there is a level-crossing between the NVs and the external spins.  Previous experiments have examined NV $T_1$ relaxation due to external paramagnetic spins often used as MRI contrast agents (e.g. Gd$^{3+}$, Mn$^{2+}$), Cu$^{2+}$, and iron ions in ferritin. The motivation is to  monitor the concentration in a microfluidic device over time \cite{T1flowimaging, ziemT1, EPRSimpson}.  Analyzing NV $T_1$ data as a function of $|\mathbf{B}_0|$ can generate the magnetic noise spectrum, identify specific paramagnetic species and yield the paramagnetic concentrations.  Further work may investigate imaging paramagnetic spins using double electron-electron resonance (DEER) \cite{marcAFMdeer}. 

\section{Conclusion and Outlook} 
In recent years, the QDM has addressed important scientific questions in diverse fields, which further motivates interest in this technology. Fortunately a QDM is relatively straightforward to build, and the technology is sufficiently mature that running a QDM experiment from start to finish is straightforward. As diamond characteristics and NV sensing techniques improve, a growing range of QDM capabilities and applications can be expected, including in extreme environments, e.g., high pressure, high temperature, and cryogenic \cite{rochAnvilCell, normAnvilCell, heziVortices}.

QDM imaging of magnetic fields is well established with a rapidly expanding application space. The sensitivity of NVs to temperature distributions and electric fields indicate that QDMs should also be applicable to imaging temperature and electric field. However, imaging a temperature inhomogeneity is challenging since temperature gradients dissipate quickly at micron length-scales in most materials, and an in-contact diamond chip will accentuate the heat dissipation from the sample being tested due to the excellent thermal conductivity of diamond, thereby, modifying the temperature profile being measured. Compared to magnetic sensing, electric sensing has the drawback that generally the electric susceptibility of a material is larger than the magnetic susceptibility, meaning that materials are often effectively transparent to magnetic fields while screening or significantly modifying electric fields. Nonetheless, QDM electric field imaging is an exciting direction that is largely unexplored. Finally, NV imaging of stress within a diamond chip (which was previously done with a single NV in an atomic force microscopy setup) is now being pursued in widefield experiments \cite{hollenbergStrain, strainMapping}. NV stress measurements can provide information about internal and external tensile and shear stress felt by the NVs, and could eventually be used to image pressure or to measure nuclear recoil tracks for particle physics experiments \cite{nvWIMPdetector}.

\section{Acknowledgements}
We acknowledge support for this work from the MITRE Corporation through the MITRE Innovation Program (MIP) and the MITRE Corporation Research Agreement No. 124787 with Harvard University. 

P.K. acknowledges support from the Sandia National Laboratories (SNL) Truman Fellowship Program,  funded by the Laboratory Directed Research and Development (LDRD) Program at SNL, operated by National Technology and Engineering Solutions of Sandia, LLC, a wholly owned subsidiary of Honeywell International, Inc., for the U.S. NNSA under contract DE-NA0003525.

\end{document}